\documentclass[final,5p,10pt,twocolumn]{elsarticle}
\usepackage[utf8]{inputenc}

\usepackage{amssymb}
\usepackage{amsmath}
\usepackage{multirow}
\usepackage{psfrag}
\usepackage{tikz}
\usetikzlibrary{arrows.meta, decorations.markings}
\usetikzlibrary{automata}
\usepackage{subcaption}

\newcommand{\EG}[0]{$\varepsilon$-greedy }
\newcommand{\ES}[0]{$\varepsilon$-sticky }

\begin{document}

\begin{frontmatter}

\title{Multi-Armed Bandits for Decentralized AP selection in Enterprise WLANs}

\author{Marc Carrascosa \corref{cor1}}
\ead{marc.carrascosa@upf.edu}
\author{Boris Bellalta}
\ead{boris.bellalta@upf.edu}
\address{Wireless Networking Research Group, Universitat Pompeu Fabra \\ Carrer de Roc Boronat 138, 08018 Barcelona, Spain}
\cortext[cor1]{Corresponding author}

\begin{abstract}

WiFi densification leads to the existence of multiple overlapping coverage areas, which allows user stations (STAs) to choose between different Access Points (APs). The standard WiFi association method makes STAs select the AP with the strongest signal, which in many cases leads to underutilization of some APs while overcrowding others. To mitigate this situation, \textit{Reinforcement Learning} techniques such as \textit{Multi-Armed Bandits} (MABs) can be used to dynamically learn the optimal mapping between APs and STAs, and so redistribute the STAs among the available APs accordingly. This is an especially challenging problem since the network response observed by a given STA depends on the behavior of the others. Therefore, it is very difficult to predict without a global view of the network.

In this paper, we focus on solving this problem in a decentralized way, where STAs independently explore the different APs inside their coverage range, and select the one that better satisfy their needs. To do it, we propose a novel approach called Opportunistic $\varepsilon$-greedy with Stickiness that halts the exploration when a suitable AP is found, only resuming the exploration after several unsatisfactory association rounds. With this approach, we reduce significantly the network response dynamics, improving the ability of the STAs to find a solution faster, as well as achieving a more efficient use of the network resources. 
  
We show that to use MABs efficiently in the considered scenario, we need to keep the exploration rate of the STAs low, as a high exploration rate leads to high variability in the network, preventing the STAs from properly learning. Moreover, we investigate how the characteristics of the scenario (position of the APs and STAs, mobility of the STAs, traffic loads, and channel allocation strategies) impact on the learning process, as well as on the achievable system performance. 

We also show that all STAs in the network improve their performance even when only a few STAs participate in the search for a better AP (i.e., implement the proposed solution). We study a case where stations arrive progressively to the system, showing that the considered approach is also suitable in such a non-stationary set-up. Finally, we compare our MABs-based approach to a load-aware AP selection mechanism, which serves us to illustrate the potential gains and drawbacks of using MABs.

\end{abstract}

\end{frontmatter}




\section{Introduction}

WiFi networks are ubiquitous nowadays, and the demand for higher data rates and area coverage keeps increasing, as well as the amount of wireless devices per user. Wired traffic accounted for 48\% of the Internet traffic in 2017, but it is expected to account for only 29\% of it by 2022, with WiFi increasing from 43\% to 51\%. This increase in the popularity of WiFi can also be seen in the number of public hotspots around the world. There were 124 million hotspots in 2017, and they are expected to increase to 549 million by 2022 \cite{cisco1}. 

Network densification by deploying more Access Points (APs) as a way of coping with the increasing traffic demands is leading to multiple overlaps between AP's coverage areas. This densification is extending to all types of deployments, from households to public spaces in cities, where in all cases multiple APs are deployed to entirely cover an area. To deal with this densification, new and future IEEE 802.11 amendments such as IEEE 802.11ax \cite{7422404}, IEEE 802.11be \cite{11bed} and IEEE 802.11ay \cite{11ay}, will offer solutions addressing specifically these kind of scenarios. In such dense network scenarios, with multiple APs available per station, selecting the best suitable one is required to both improve individual and  collective performance. 


The standard association for IEEE 802.11 networks uses the Strongest Signal (SS) method to associate a user station (STA) to an AP. It scans the spectrum for all possible available networks and chooses the one with the highest Received Signal Strength Indicator (RSSI) from the received beacons. This method can lead to uneven loads by overcrowding a single AP and leaving others underused~\cite{Balachandran:2002:CUB:511334.511359}, thus Enterprise WLANs that consist of multiple APs are in need of new association schemes that leverage such a situation, distributing the STAs among the available APs in a way that maximizes the quality of the users experience. 
 There have been numerous proposals to improve AP selection by STAs, also called user or station association (see Section \ref{RelWork}, \textit{Related Work}). These proposals are generally designed to estimate the free capacity of each AP within their coverage area, by using the round trip delay of probe packets, or by listening to the channel to estimate how much airtime is available. Then, they typically select the AP that is expected to offer the best service. While all those proposals may work efficiently in most cases and scenarios, they may also fail when confronted with unexpected situations, for which they were not designed. For instance, in dense networks, with many STAs, these algorithms may show a ping-pong effect and never converge to a good solution due to the rigidity of their pre-programmed sequence of actions.

In this work, our aim is to evaluate the suitability of using Reinforcement Learning to improve the performance of Enterprise WLANs by finding a feasible AP-STA association, and evaluate if these techniques can cope with the aforementioned issues by simply implementing general exploration-exploitation policies. In particular, we model the AP-STA association problem as a Multi Armed Bandits (MAB) problem, in which an agent placed in each STA can take multiple actions (i.e., choose an AP from the set of available APs). The goal of the agent is to find a way to maximize its rewards by exploring the set of available actions, learning more about the network at each step, and exploiting the most suitable alternative previously found.

In such a scenario, the main challenge to solve is that the reward obtained for each action depends on the actions taken by the other STAs, which operate independently, and thus, choosing the same action at different time instants may result in different rewards, significantly increasing the action's uncertainty. To mitigate this effect, we introduce the Opportunistic $\varepsilon$-greedy algorithm with Stickiness. It follows the default exploration-exploitation tradeoff of the basic $\varepsilon$-greedy algorithm, but it includes two other features: 1) It is opportunistic in the sense that it halts the exploration when it finds a satisfactory AP, and 2) when an AP becomes unsatisfactory, instead of exploring other APs immediately, STAs stick to it for several consecutive unsatisfactory association rounds before exploring other APs. This approach aims to improve the convergence time by reducing the number of STAs doing the exploration at the same time, and mitigate the influence of the other STAs in the learning.

Therefore, the main contributions of this paper are:
\begin{itemize}
    \item The  know-how obtained by using the MABs framework to model the association problem in Enterprise WLANs. We are convinced such know-how can be also applied to many other problems, even from different fields.
    \item A new opportunistic MAB algorithm based on \EG that includes the concept of stickiness, allowing STAs that are satisfied to remain associated to the current AP unless they become unsatisfied for several consecutive association rounds. It is based on the concept of halting the exploration when a feasible solution is found, even if it is not the optimal one, so it contributes to reduce the non-stationarity of the multi-agent environment. 
    \item The characterization of the behavior of the new MAB algorithm presented, showing how it provides a much better performance than $\varepsilon$-greedy. A key aspect of \ES is that it converges faster to a feasible solution, if it exists, at the price of giving up further exploration to find the optimal one.
    \item A detailed performance evaluation, giving insights on the effect that STA placement, the number of orthogonal channels, the channel width and variable loads have on the performance of the algorithm. 
    
    \item We analyze the behavior of the network when a limited amount of STAs implement the MAB framework. We study  how these algorithms cope with the progressive arrival of STAs to the network, and the effects of user mobility as well. Lastly, we compare  our framework to a load-aware AP selection mechanism.
\end{itemize}


Beyond the paper's contributions, it also offers some general lessons on the use of MABs to solve the AP selection problem:
\begin{enumerate}
    \item Characteristics of the scenario: MABs work for all AP and STA deployments, but excel when the APs are in a grid and the STAs are placed non-uniformly. These situations create particular and unbalanced situations where the \textit{by default} solution to the problem does not work properly, and so a new one must be learnt.
    \item Scenario non-stationarity and exploration level: variable traffic loads, user mobility and a progressive arrival of STAs to the network  increase the variability of the network. As expected, the higher the variance is, the harder learning becomes, and lower are the benefits of using MABs. In those scenarios, combining an  aggressive exploration phase when changes are detected with a  mechanism to halt the exploration if a feasible solution is found seems appropriate.
    \item History and Past information: As the reward is updated every round, using the average of all rounds is enough to compensate for network changes. A window can be used to consider variability, but it needs to be  large enough to get proper values, but short enough to avoid covering periods with different statistics.
    \item  Optimal vs feasible solutions:We observe how feasible solutions ( i.e., those that improve the satisfaction of users in our case) can be reached relatively fast. However, optimal solutions (i.e., the best feasible solution) are hard to be found in a decentralized way.
\end{enumerate}

The rest of the paper is structured as follows. Section \ref{MABs}\\ gives an introduction to the Multi-Armed Bandit problem. Section \ref{toyschapter} introduces the Enterprise WLAN scenario, providing also an illustrative example of the AP selection problem. In Section \ref{MABs2} the MAB algorithms used in this paper are described. Section \ref{perfEval} contains the results obtained in a simulated environment  comparing \EG and $\varepsilon$-sticky, and Section \ref{perfEval2} further tests our MAB framework with more challenging problems. The related work can be found in Section \ref{RelWork}. A final summary can be found in Section \ref{concl}, as well as several future research directions.

This paper significantly extends our previous work in~\cite{Carr1906:Decentralized}. We have included a new section giving some background on MABs, as well as an illustrative toy scenario further detailing the association problem. Moreover, we have also updated the way we calculate the reward received by each agent, and have considered the case in which APs are deployed in a grid. The results section has been enhanced to include the trade-off of using wider channels at the cost of decreasing the number of available orthogonal channels. We also address the effects of variable traffic load requirements, the case in which only a fraction of the STAs implement the \EG agents, the progressive arrival of STAs to the network, and the performance under user mobility. These additions result in more randomness, which increases the complexity of the problem studied.  Further, a section considering another AP selection method based on rules instead of MABs has been added.


\section{Multi-armed Bandits}\label{MABs}

The Multi-Armed Bandit problem models a scenario in which a learning agent has to choose between $\kappa$ actions, often called arms, over time in rounds. For each action taken, the agent receives a reward $\mu$ from the environment. Finding the optimal arm becomes an exploration-exploitation trade-off in which the agent has to decide between exploring arms to obtain information about the environment, or selecting the arm that has historically given the highest reward (exploitation). The way this decision is made is the focus of the MAB algorithms. Ultimately, the objective of the agent is to maximize the long term reward, usually also managing a trade-off between the rate of learning and reaching optimal results (i.e., a fast learning rate may lead to the algorithm not exploring enough options and finding a suboptimal solution, while a slow learning rate may waste time taking redundant actions). The name for this problem comes from a classic example in which a gambler plays several slot machines in a casino at the same time. An in-depth introduction to MABs can be found in \cite{introtoMABs}.

MABs can be classified into different types:
\begin{itemize}
    \item In \textbf{Stochastic bandits} \cite{robbins1952some},  actions have independent and identically distributed (iid) rewards. Each arm follows its own distribution of rewards, and the algorithm only gets a reward for the action taken. Common algorithms used for these MABs are \EG and UCB \cite{Auer2002}. A common example is the previously mentioned casino with slot machines, each machine having a different payout distribution and an agent trying to find the machine that will give a prize more often.
    
    \item In \textbf{Bayesian bandits} the exploration is modeled according to the rewards received, so that the probability of choosing arm $k_i$ is proportional to the rewards obtained historically from it. Thompson sampling \cite{10.2307/2332286, Russo:2018:TTS:3283245.3283246} is a popular implementation that is also commonly used for stochastic bandits. 
    
    \item \textbf{Adversarial bandits }forego the assumption of arms following a fixed distribution, instead considering that an opponent has control of the payoffs. Algorithms related to this type of bandits are Hedge and EXP3 \cite{adv}, which give weights to the arms and update these weights according to the rewards received, then actions are taken according to the weight distribution.
    
    \item \textbf{Non-stationary bandits} are a version of stochastic bandits in which the number of players, arms, and the reward distribution can change over time \cite{besbes2014stochastic}. These bandits find themselves in between the stochastic and adversarial bandits.

    \item In \textbf{Contextual bandits} the agent also has access to some information (or context) from the environment that relates to the payoffs in the next round \cite{slivkins2014contextual}. Commonly associated with the problem of placing advertisements in web pages, using the context of web features and user profiling, where the intent is to maximize the probability of users clicking on ads.

\end{itemize}

One way to evaluate the performance of an algorithm is the regret. It compares the reward obtained over $T$ rounds to the reward $\mu^*$ achievable by always taking the optimal arm. The regret is defined as: 
	\begin{equation}
    \begin{split}
		R(T) = \mu^* \cdot T - \sum_{t=1}^T \mu(k_t)
	\end{split}
\end{equation}

An algorithm will perform successfully if the expected regret $\mathbb{E}[R(T)]$ decreases over time until it finally converges to 0, meaning that the agent is capable of learning the action with the highest payoff.

\section{Enterprise WLANs: System Model} \label{toyschapter}

\subsection{Path-loss and transmission rate selection}

The path-loss (i.e., signal attenuation due to propagation) between the APs and the STAs is obtained using the 5 GHz TMB model for indoors scenarios \cite{tmb} given by:
	\begin{equation}
    \begin{split}
		\text{PL}_{\text{TMB}}(d_{i,j})= L_0 + 10\gamma\log_{10}(d_{i,j}) + k\overline{W} d_{i,j} + G_s
	\end{split}
\end{equation}
where $d_{i,j}$ is the distance between STA $i$ and AP $j$, $L_0$ is the path loss at one meter, and $\gamma$ is the path-loss exponent, $k$ is the wall attenuation factor, and $\overline{W}$ is the average number of traversed walls per meter. $\text{G}_s$ is a random variable uniformly distributed with mean 5 modelling the shadowing. For all those parameters, the same values as in \cite{tmb} are used, which can also be found in Table \ref{teVar}.	

Using the obtained $\text{PL}_{\text{TMB}}$ values, we obtain the received power used for the communication between each AP-STA pair, i.e. $P_r = P_t -\text{PL}_{\text{TMB}}(d_{i,j})$. Then, using the received power as a reference, we obtain both the transmission rate, $r$, and the legacy transmission rate $r_L$.
\setlength\tabcolsep{3 pt}
	\begin{table}[ht]\centering
	\begin{small}
    \begin{tabular}{|p{4cm}|c|c|}
  		\hline  
		\textbf{Name}  &   \textbf{Variable}& \textbf{Value}\\ \hline
		Legacy preamble  & $T_{\text{PHY-legacy}}$ & $20 \mu s$\\ \hline
		HE Single-user preamble & $T_{\text{PHY-HE-SU}}$ & $52 \mu s$\\ \hline
		OFDM symbol duration& $\sigma$ & $16 \mu s$\\ \hline
		OFDM Legacy symbol dur. & $\sigma_{\text{Legacy}}$ & $4 \mu s$\\ \hline
        Short InterFrame Space & SIFS & $16 \mu s$\\ \hline
        DCF InterFrame Space & DIFS & $34\mu s$\\ \hline
		Average back-off duration  & $E[\psi]$ & 7.5 slots\\         \hline
        Empty backoff slot & $T_e$ & $9 \mu s$\\ \hline \hline
		Service Field   &  $L_{\text{SF}}$ & 32 bits\\ \hline
		MAC header   & $L_{\text{MH}}$ & 272 bits\\ \hline
		Tail bits & $L_{\text{TB}}$& 6 bits\\  \hline
		ACK bits & $L_{\text{ACK}}$ & 112 bits \\ \hline
		Frame size &  $L$ & 12000 bits\\  
		\hline \hline
		Path loss intercept  & $L_0$ & 54.1200 dB \\ \hline
		Path loss exponent & $\gamma$ & 2.06067 \\\hline
		Wall attenuation & $k$ & 5.25 dB\\\hline
		
	\end{tabular}
	\caption{Notation, simulation parameters and values. }
	\label{teVar}
	\end{small}
	\end{table}


\subsubsection{Airtime and throughput calculation}\label{appref}

The required airtime  (i.e., the fraction of time required for the transmission of the data) per STA and per second is calculated taking into account the throughput required by a station, $w$, the average packet sizes it transmits, $L$, the transmission rate $r$, and all other IEEE 802.11 overheads. The notation and values used can be found in Table \ref{teVar}. 

The required airtime for any STA $i$ is given by:
\begin{align} \label{eq:flow_util}
	u_i(\omega_i,L,r_i,r_{L,i})=\frac{w_i}{L} \cdot (E[\psi] T_e + T(L,r_i,r_{L,i}))
\end{align}
where the calculation of $T(L,r_i,r_{L,i})$ can be found in  \ref{appA}.

AP $j$ observes the following channel occupancy :
\begin{align} \label{eq:ap_util}
	U_{j}=\min(1,\sum_{\forall i \in \mathcal{S}_j}{u_i(\omega_i,L,r_i,r_{L,i})})
\end{align}
where $\mathcal{S}_j$ is the set of all stations associated to AP $j$ and to other APs within the coverage range of AP $j$ that operate in the same channel.

\begin{figure}[ht]
    \centering
    \includegraphics[width=0.45\textwidth]{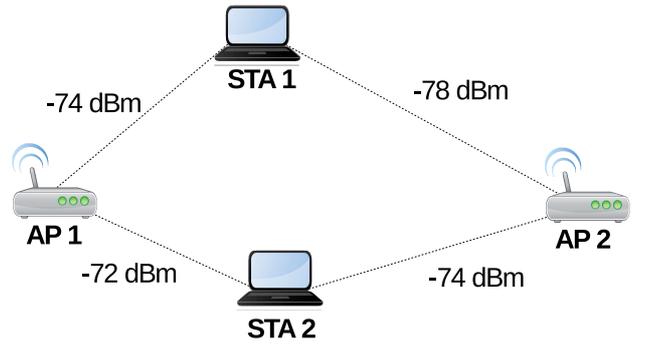}
    \caption{Toy Scenario. }
    \label{toyS}
\end{figure}
Finally, the throughput received by STA $i$ is obtained from:
\begin{align} \label{eq:thr}
\zeta_{i}&=u_i'\frac{L}{E[\psi] T_e + T(L,r_i,r_{L,i})} \nonumber\\ 
&=\frac{u_i(\omega_i,L,r_i,r_{L,i})}{U_j} \frac{\omega_i}{u_i(\omega_i,L,r_i,r_{L,i})} = \frac{\omega_i}{U_j}
\end{align}
where $u_i'=\frac{u_i(\omega_i,L,r_i,r_{L,i})}{U_j}$.

\subsection{The association anomaly in WiFi: a toy example}

The default association method for 802.11 devices is to scan for all nearby APs and select the one with the strongest RSSI. This method can lead to an uneven network distribution where resources go unused and STAs starve. Such a scenario is shown in Figure \ref{toyS}, where we have two APs and two STAs. STA 1 requires 12 Mbps of throughput, while STA 2 requests 15 Mbps. The APs are in orthogonal channels.

 Table \ref{rssivsopt} shows all possible STA associations in this particular scenario, with the necessary airtime  for the current connection, as well as the achievable throughput. Since both STAs are closer to AP 1 than AP 2, the default Strongest Signal (SS) method will associate both STAs to AP 1. As the required airtime for both connections exceeds the capacity of AP 1, each of them will receive less throughput than they desire. The same happens if both STAs connect to AP 2.

\setlength\tabcolsep{5 pt}
	\begin{table}[ht]\centering

	\begin{small}
    \begin{tabular}{|l|c|c|c|c|}
  		\hline  
  		 
		\textbf{Association}& \textbf{Throughput}&\textbf{Airtime}\\ \hline
	    STA 1 to AP 1 & 7.6 Mbps  & 0.4951/0.7825 (63.27\%) \\
	    STA 2 to AP 1  & 9.5 Mbps  & 0.5049/0.7981  (63.26\%)  \\\hline
	   \textbf{STA 1 to AP 1}  & \textbf{12 Mbps}  & 0.7825/0.7825 (100\%) \\
	    \textbf{STA 2 to AP 2}  & \textbf{15 Mbps}  &  0.9781/0.9781 (100\%) \\\hline
		STA 1 to AP 2 & 11.3 Mbps  & 1/1.0585 (94.47\%)\\
	    STA 2 to AP 1  & 15 Mbps  & 0.7981/0.7981 (100\%)\\\hline
        STA 1 to AP 2  & 5.9 Mbps  & 0.5197/1.0585 (49.09\%)\\
	    STA 2 to AP 2  & 7.4 Mbps  & 0.4803/0.9781 (49.10\%)\\\hline
	\end{tabular}
		\caption{Possible association combinations and their throughput and airtime.}
	\label{rssivsopt}
	\end{small}
	\end{table}

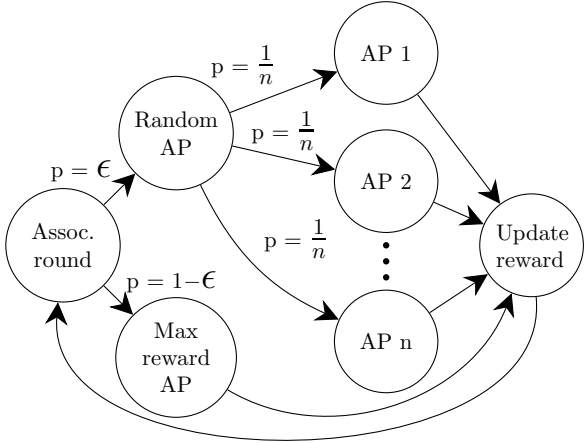
\begin{figure}[h]
    \centering

\begin{tikzpicture}[scale = 0.85, transform shape,-> ,auto,-{Stealth[ ,length=3mm,width=3mm]}]
 \node [ state ,align=center, text=black,minimum width=1.6 cm,  text width=1.3cm](A) at (0,0){Assoc. round};  
 \node [ state ,align=center,text=black,minimum size=1.6 cm,  text width=1.3cm](B) at (1.75,1.75){Random AP};  
 \node [ state ,align=center,text=black,minimum size=1.6 cm,  text width=1.1cm](C) at (1.75,-1.75){Max reward AP};
 \node [ state ,text=black,minimum size=1.6 cm](D) at (5,3){AP 1};
 \node [ state ,text=black,minimum size=1.6 cm](E) at (5,1){AP 2};
  \node [ state ,text=black,minimum size=1.6 cm](F) at (5,-1.5){AP n};
  
  
  \node [ state ,text=black,minimum size=1.6 cm, text width=1.1cm](G) at (7.25,0){Update reward};
\node[circle,fill,inner sep=1pt](d1) at (5,0){};
\node[circle,fill,inner sep=1pt](d2) at (5,-0.5){};
\node[circle,fill,inner sep=1pt](d3) at (5,-0.25){};

 \path [text=black](A) edge              node {p = \LARGE{$\epsilon$}} (B);
 \path [text=black, above right, midway ](A) edge              node {p = $1 - $\LARGE{$\epsilon$}} (C);
 \path [text=black](B) edge              node {p = \Large{$\frac{1}{n}$}} (D);
 \path [text=black, above](B) edge              node {p = \Large{$\frac{1}{n}$}} (E);
  \path [text=black ](B) edge[bend right=20]  node {p = \Large{$\frac{1}{n}$}} (F);
  \path [text=black ](D) edge  node {} (G);
 \path [text=black ](E) edge  node {} (G);
  \path [text=black ](F) edge  node {} (G);
   \path [text=black ](C) edge[bend right=50]  node {} (G);
   \path [text=black ](G) edge[bend left=95]  node {} (A);
\end{tikzpicture}

\caption{$\varepsilon$-greedy state diagram. }
    \label{e-g}
\end{figure}

 The connection between STA 1 and AP 2 is bad enough that STA 1 by itself needs more airtime than AP 2 can afford. This means that STA 1 can only achieve its full throughput on AP 1. STA 2 has a worse connection with AP 2 than AP 1, but it can still fulfill its needs on it. Thus, the optimal association is the association that minimizes the load of each AP. For this particular scenario, STA 1 should associate to AP 1 and STA 2 to AP 2.


\section{MAB-based AP-selection mechanisms}\label{MABs2}

\subsection{\EG}

Each STA has an agent implementing the \EG algorithm, and each of the APs in the sensing range of the STA is an arm. The agent keeps track of the reward for each arm available. The value of $\varepsilon$ modifies the exploration rate. Each exploration or exploitation may lead to a reassociation if the AP selected is different from the current one. The \EG behavior is shown in Figure \ref{e-g}.

The \EG algorithm works in iterations of time $t$ which we will call association rounds. For our implementation we need to consider enough time for a STA to associate to a new AP and then perform a download that allows us to measure the link capacity. We consider $t=180 s$ to be enough for our purposes. We use the normalized average throughput as the reward for each AP.

\begin{figure}[ht]
\centering
        \includegraphics[width=.5\textwidth]{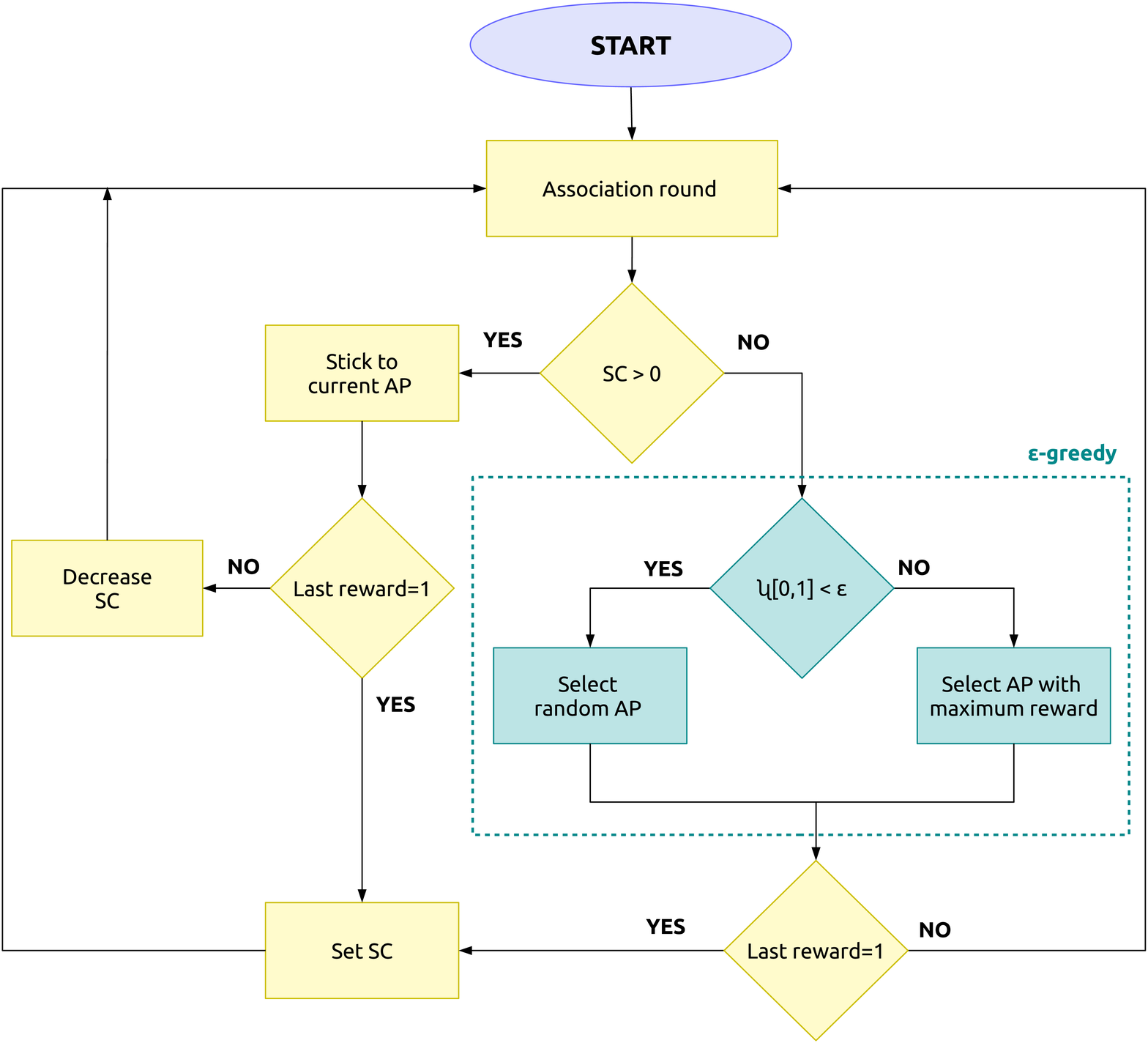}
    \caption{Block diagram for $\varepsilon$-sticky. }
    \label{stickyblocks}
\end{figure}

\subsection{\ES or opportunistic \EG with stickiness}

We extend the \EG algorithm by including the concept of stickiness, i.e., not taking any actions for a period of time after finding a favorable situation. If a STA finds an AP that can provide all its requested throughput, then the STA will hold the association for SC (Sticky Counter) consecutive rounds and will only explore again if the received throughput is insufficient for all those rounds. This method avoids needless exploration when there are small temporary changes in the network. Figure \ref{stickyblocks} details the \ES algorithm.

\subsection{Reward calculation}

The reward of AP $j$ at association round $k$ is calculated as the average of the rewards received by AP $j$ in current and previous association rounds, i.e.,

\begin{align}
     R_{\text{AP}_j} = \frac{1}{N_{\text{AP}_j}}\sum_{i=1}^{N_{\text{AP}_j}}{\zeta_i}  
\end{align}
 where $N_{\text{AP}_j}$  is the number of times AP $j$ has been selected. 

\subsection{Example}

\setlength\tabcolsep{8 pt}
	\begin{table*}[ht]\centering

	\begin{small}
    \begin{tabular}{|l|c|c|c|c|c|c|c|c|c|c|c|c|c|c|}
  		\hline  
  	      \multicolumn{2}{|c|}{} & \multicolumn{12}{|c|}{\textbf{Reward}}\\\hline
		\multicolumn{2}{|c|}{\textbf{Assoc. round}} & \textbf{1}& \textbf{2}& \textbf{3}& \textbf{4}& \textbf{5}& \textbf{6}& \textbf{7}& \textbf{8}& \textbf{9}&\textbf{10}& \textbf{11}& \textbf{12}\\ \hline
	    \multirow{2}{*}{\textbf{STA 1}}& \textbf{AP 1} & $\mathbf{0.63}$ & $\mathbf{0.81}$ & $0.81$ & $\mathbf{0.87}$ & $\mathbf{0.81}$ & $\mathbf{0.85}$ & $\mathbf{0.87}$ & $\mathbf{0.89}$ & $\mathbf{0.90}$ & $\mathbf{0.91}$ & $\mathbf{0.92}$ & $\mathbf{0.89}$ \\\cline{2-14}
	    & \textbf{AP 2} & $0$ & $0$ & $\mathbf{0.49}$ & $0.49$& $0.49$& $0.49$& $0.49$& $0.49$& $0.49$& $0.49$& $0.49$& $0.49$\\\hline\hline
	    \multirow{2}{*}{\textbf{STA 2}}& \textbf{AP 1} & $\mathbf{0.63}$ & $0.63$& $0.63$& $0.63$& $\mathbf{0.63}$& $0.63$& $0.63$& $0.63$& $0.63$& $0.63$& $0.63$& $\mathbf{0.63}$  \\\cline{2-14}
	    & \textbf{AP 2} & $0$ & $\mathbf{1}$ & $\mathbf{0.74}$ & $\mathbf{0.83}$& $0.83$& $\mathbf{0.87}$& $\mathbf{0.89}$& $\mathbf{0.91}$& $\mathbf{0.92}$& $\mathbf{0.93}$& $\mathbf{0.94}$& $0.94$\\\hline
	    
	\end{tabular}
		\caption{Average reward over time for $\varepsilon$-greedy. Bolded rewards show the chosen AP for each association round. }
	\label{exampleEG}
	\end{small}
	\end{table*}

\setlength\tabcolsep{8 pt}
	\begin{table*}[ht]\centering

	\begin{small}
    \begin{tabular}{|l|c|c|c|c|c|c|c|c|c|c|c|c|c|c|}
  		\hline  
  		 \multicolumn{14}{|c|}{\textbf{Case 1}}\\\hline
  	      \multicolumn{2}{|c|}{} & \multicolumn{12}{|c|}{\textbf{Reward}}\\\hline
		\multicolumn{2}{|c|}{\textbf{Assoc. round}} & \textbf{1}& \textbf{2}& \textbf{3}& \textbf{4}& \textbf{5}& \textbf{6}& \textbf{7}& \textbf{8}& \textbf{9}&\textbf{10}& \textbf{11}& \textbf{12}\\ \hline
	    \multirow{2}{*}{\textbf{STA 1}}& \textbf{AP 1} & $\mathbf{0.63}$ & $0.63$& $0.63$& $0.63$& $\mathbf{0.63}$& $\mathbf{0.63}$& $0.63$& $\mathbf{0.72}$& $\mathbf{0.77}$& $\mathbf{0.81}$& $\mathbf{0.84}$& $\mathbf{0.86}$  \\\cline{2-14}
	    & \textbf{AP 2} & $0$ & $\mathbf{0.94}$ & $\mathbf{0.94}$ & $\mathbf{0.94}$& $0.94$& $0.94$& $\mathbf{0.82}$& $0.82$& $0.82$& $0.82$& $0.82$& $0.82$\\\hline\hline
	     \multirow{2}{*}{\textbf{STA 2}}& \textbf{AP 1} & $\mathbf{0.63}$ & $\mathbf{0.81}$ & $\mathbf{0.87}$ & $\mathbf{0.90}$ & $\mathbf{0.85}$ & $\mathbf{0.81}$ & $0.81$ & $0.81$ & $0.81$ & $0.81$ &$0.81$ & $0.81$ \\\cline{2-14}
	    & \textbf{AP 2} & $0$ & $0$ &$0$ & $0$ & $0$& $0$ & $\mathbf{0.49}$ & $\mathbf{0.74}$ & $\mathbf{0.83}$ & $\mathbf{0.87}$ & $\mathbf{0.89}$ & $\mathbf{0.91}$ \\\hline
	      		 \multicolumn{14}{|c|}{\textbf{Case 2}}\\\hline
         \multirow{2}{*}{\textbf{STA 1}}& \textbf{AP 1} &$\mathbf{0.63}$& $\mathbf{0.81}$& $\mathbf{0.87}$& $\mathbf{0.90}$& $\mathbf{0.92}$& $\mathbf{0.93}$& $\mathbf{0.94}$& $\mathbf{0.95}$& $\mathbf{0.95}$& $\mathbf{0.96}$& $\mathbf{0.96}$& $\mathbf{0.96}$\\\cline{2-14}
	    & \textbf{AP 2} & $0$ & $0$& $0$& $0$& $0$& $0$& $0$& $0$& $0$& $0$& $0$& $0$\\\hline\hline
	     \multirow{2}{*}{\textbf{STA 2}}& \textbf{AP 1} & $\mathbf{0.63}$  & $0.63$& $0.63$& $0.63$& $0.63$& $0.63$& $0.63$& $0.63$& $0.63$& $0.63$& $0.63$& $0.63$\\\cline{2-14}
	   & \textbf{AP 2} & $0$ & $\mathbf{1}$ & $\mathbf{1}$& $\mathbf{1}$& $\mathbf{1}$& $\mathbf{1}$& $\mathbf{1}$& $\mathbf{1}$& $\mathbf{1}$& $\mathbf{1}$& $\mathbf{1}$& $\mathbf{1}$\\\hline
	\end{tabular}
		\caption{Average reward over time for $\varepsilon$-sticky. Bolded rewards represent current association.}
	\label{exampleES}
	\end{small}
	\end{table*}
	
We will show how  both \EG and \ES perform following the toy scenario introduced in Figure~\ref{toyS}. In this example we will use  $\varepsilon = 0.3$ and a sticky counter of 2. We will follow the algorithms for 12 association rounds.

We start with $\varepsilon$-greedy. The reward at each step can be found in Table \ref{exampleEG}, and the AP selected by each STA is bolded for each round. 

We use SS for the first association and find that both STAs associate to AP 1 and receive a reward of $0.63$. In round 2, STA 1 exploits and picks the AP with maximum reward, in this case AP 1. STA 2 explores and tries AP 2. In this case both of them receive a higher reward than before, $0.94$ for STA 1 in AP 1(increasing its AP 1 average reward to $0.81$) and $1$ for STA 2 in AP 2. In round 3 both STAs explore and associate to AP 2. As this is the worst association possible, STA 1 gets a low reward for AP 2, and STA 2 lowers its average reward for AP 2. In round 4 they both exploit, STA 1 associating to AP 1 and STA 2 to AP 2. This is the optimal solution as shown in Table \ref{rssivsopt}, so they both get a 1 for their reward, raising their respective averages for the corresponding APs. Both STAs  explore in round 5 to AP 1. Then, in round 6 they exploit again to the optimal association. In round 7, STA 1 explores but due to the randomness of exploration, it stays in AP 1, the best option according to its rewards. In rounds 8, 9 and 10, the optimal configuration is kept until round 12 in which STA 2 explores to AP 1 again. We can observe however that at this point STA 1 has a higher reward for AP 1, and STA 2 for AP 2, so while they may explore 30\% of the time, they will choose the optimal configuration 70\% of the time. 

Let us now consider $\varepsilon$-sticky in Table \ref{exampleES}, in which we show two different solutions. In the first case, both STAs start with SS and associate to AP 1. As none of them are fully satisfied, they do not stick to their AP and use the \EG rules. In round 2, STA 1 explores AP 2, finding a reward of $0.94$, while STA 2 exploits and finds a reward of 1 for AP 1, and so it will stick to this AP in the following round. Since STA 1 has found a higher reward in AP 2, it now exploits AP 2 in round 3 and 4, while STA 2 continues to stick to AP 1. In round 5 STA 1 explores to AP 1, which makes both of them unsatisfied. The sticky counter of STA 2 decreases by 1. In round 6 STA 1 explores to AP 1 again, and now the sticky counter of STA 2 is decreased to 0, meaning that it will use \EG for the next round. In round 7 STA 1 exploits to AP 2 and STA 2 explores AP 2, both receiving low rewards. In round 8 STA 1 explores to AP 1 and STA 2 explores to AP 2, both finding a reward of 1 and making them stick to each AP for the remainder of the rounds. 
In case 1 we show that stickiness may reach a situation in which a satisfied STA delays the network in converging to the optimal solution, with the exploration of STA 1 in the second round leading to the second best configuration, and both STAs repeating it due to the stickiness of STA 2 and the higher reward for STA 1. It does not completely block the network from learning the optimal situation however, as we only require the exploration of STA 1 to stop STA 2 from sticking by exploring as many times as the value of the sticky counter. From this we can infer that the value of the sticky counter should be kept low, or at least that the value of $\varepsilon$ should increase with it. Note however, that for all the association rounds that were spent in the second best configuration, both STAs received a higher throughput than with the SS method.

In case 2 we show the optimal scenario, in which STA 2 explores AP 2 in round 2, and then both STAs stick to their APs as they have both found a satisfactory association, finding the optimal solution with a single reassociation.

Finally, these two cases further show that \ES does not make any distinction between two viable configurations, as it will stick to any AP capable of providing all the requested load.

\begin{figure*}[ht]
\centering
   \begin{subfigure}[b]{.4\textwidth}
        \includegraphics[width=\textwidth]{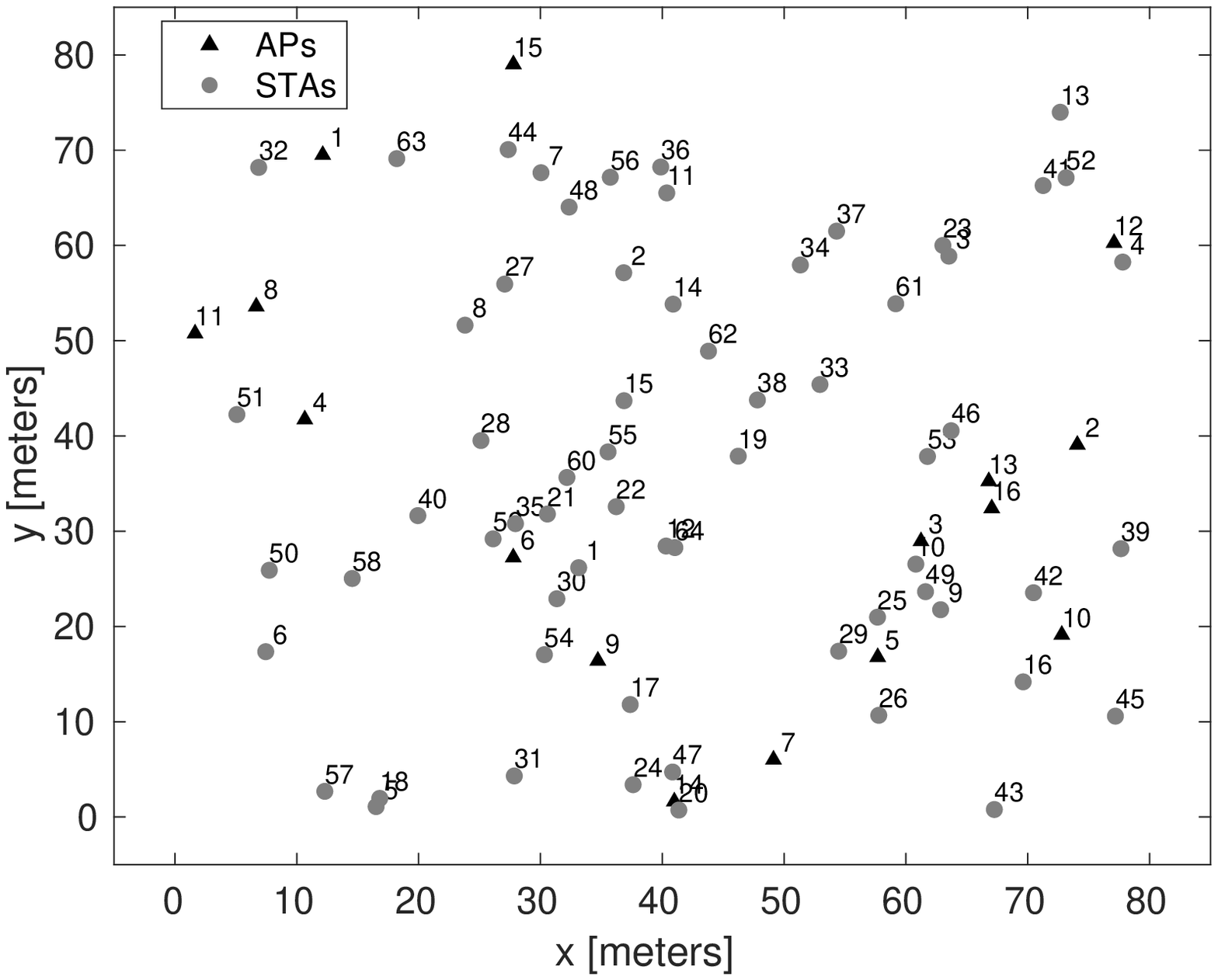}
    \caption{AP and STAs distributed uniformly at random.}
    \label{scenRand}
    \end{subfigure}
    \begin{subfigure}[b]{.4\textwidth}
        \includegraphics[width=\textwidth]{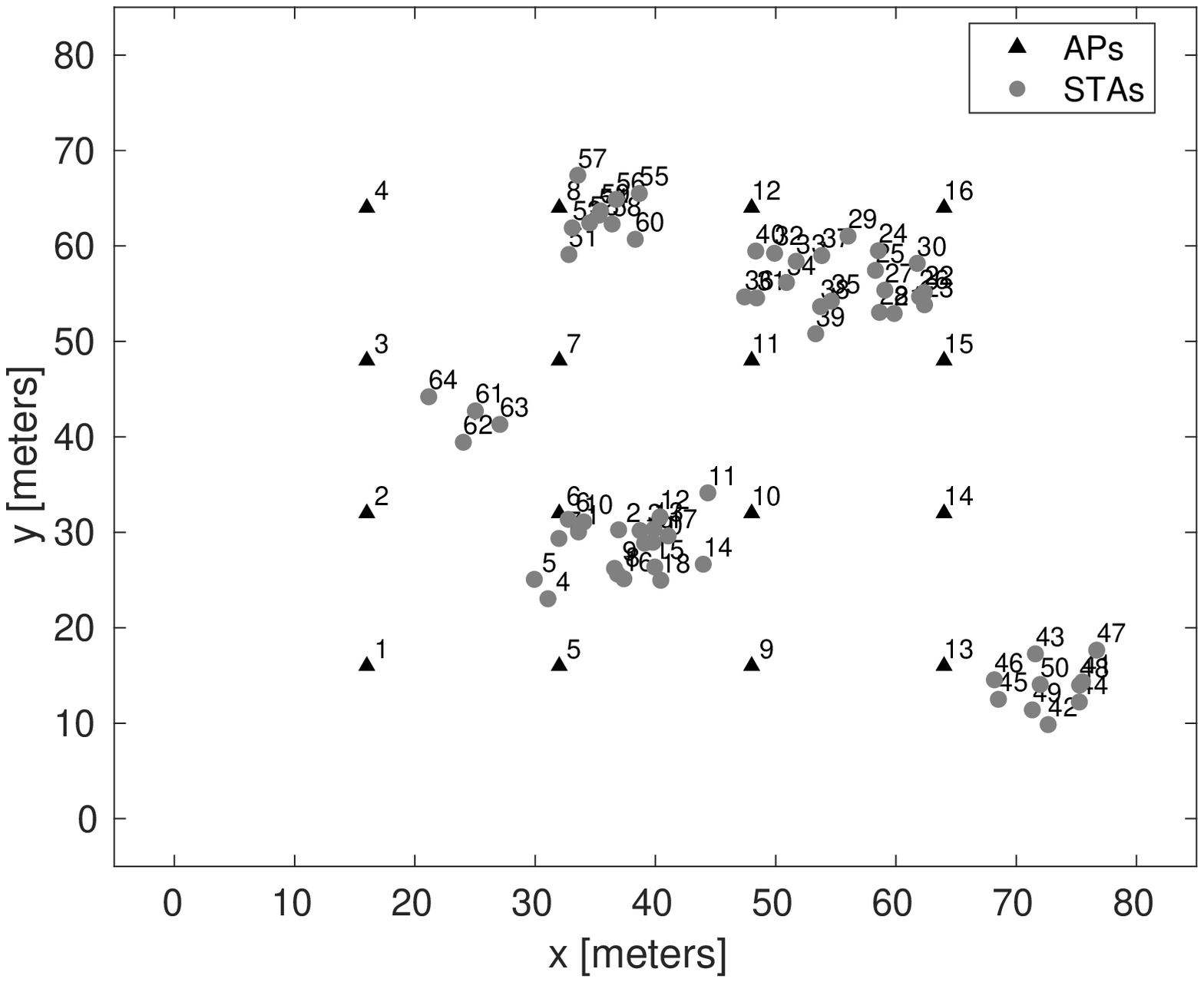}
    \caption{APs in a grid with clustered STAs.}
    \label{scenClust}
    \end{subfigure}
     \caption{ Snapshot of the scenarios considered in this paper.}
     \label{Fig:Scenarios}
\end{figure*}

\subsection{Note about implementation complexity in practice}

Although the implementation and evaluation of the proposed framework in practice is out of the scope of this work, we would like to note that it would require a simple modification in the firmware of the end user devices. After the STA does a full scan it would need to store the MAC addresses of all present APs and their associated reward. Moreover, the computational complexity of the algorithms is also low, only requiring to use a random number generator. Each AP reassociation would use the frames as defined by the IEEE 802.11 standard. To take into account STAs' mobility, it would also be enough to save multiple AP tables, one for each list of APs found in the full scan (i.e., a table for all frequently visited locations from the user), so that information would not be lost after the user leaves an area.


\section{Understanding the gains of \ES vs \EG}\label{perfEval}

In this Section, we aim to understand the reasons why \ES outperforms $\varepsilon$-greedy. To do that, we consider different scenarios, different approaches to calculate the reward, and test different possible configurations for both MAB schemes.
\begin{figure*}[h]
\centering
   \begin{subfigure}[b]{.4\textwidth}
        \includegraphics[width=\textwidth]{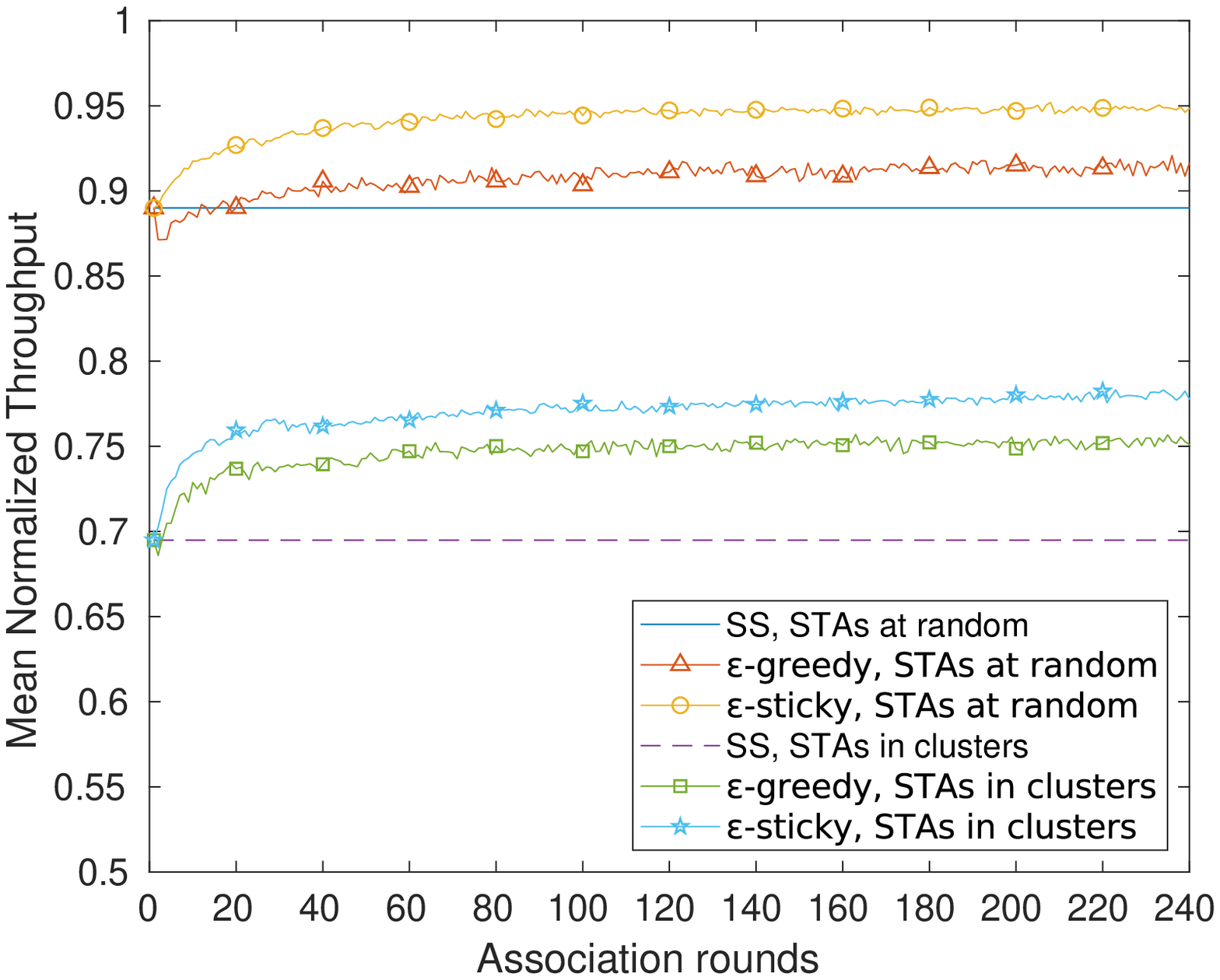}
    \caption{APs are distributed uniformly at random.}
    \label{randbox}
    \end{subfigure}
    \begin{subfigure}[b]{.4\textwidth}
        \includegraphics[width=\textwidth]{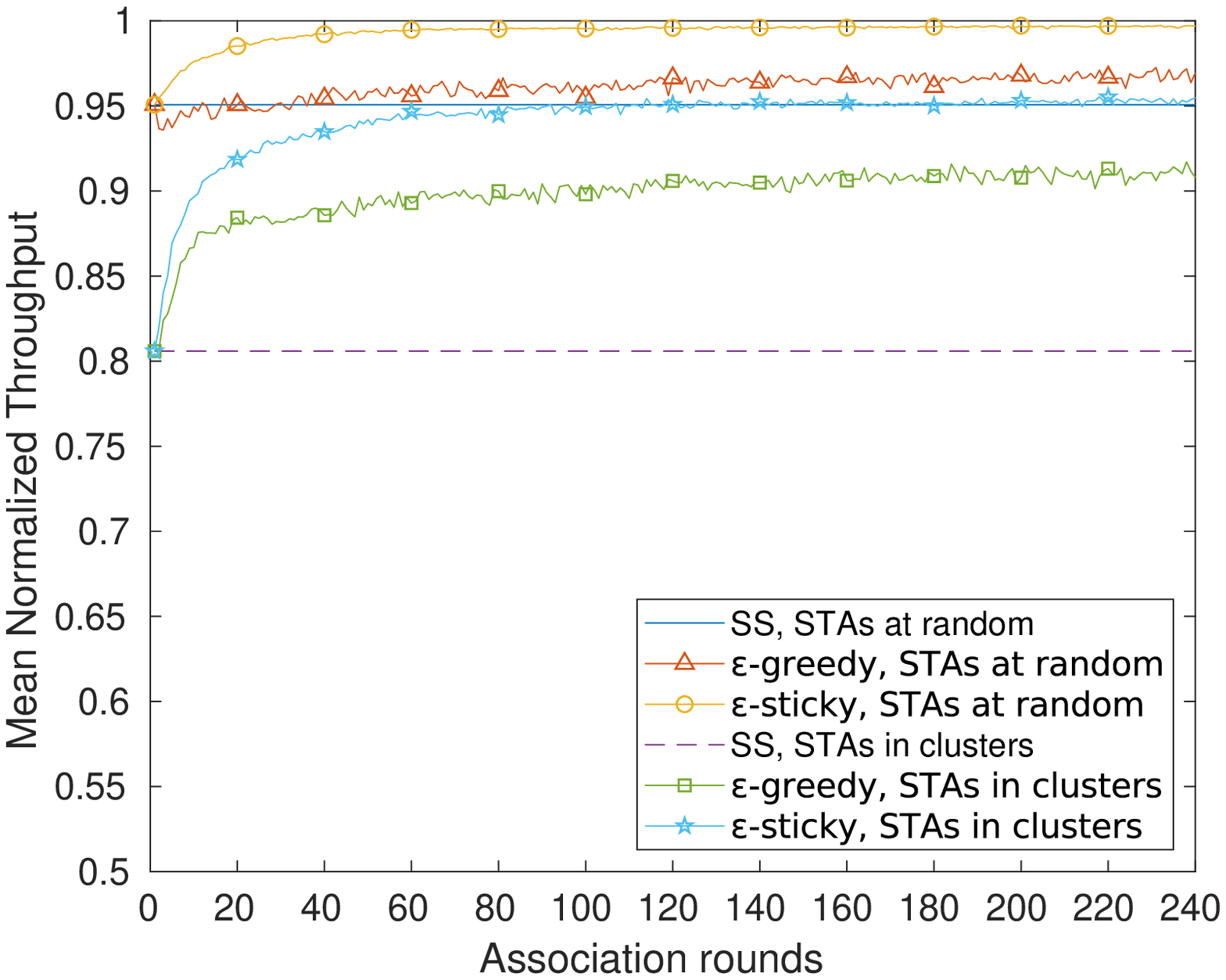}
    \caption{APs are placed in a grid.}
    \label{gridbox}
    \end{subfigure}
     \caption{Mean throughput obtained in different network deployments with \EG and $\varepsilon$ -sticky.}
     \label{Fig:Scenarioss}
\end{figure*}
\subsection{Simulation set-up}

In our simulations we use an area of $80 \times 80$ metres with $M$ APs and $N$ STAs. APs can be placed in a grid or uniformly at random. STAs can also be placed uniformly at random or in small clusters of $N_c$ STAs in areas of $10 \times 10$ m. The throughput in Mbps requested by each STA can be fixed or taken at random from a given set of values. Each simulation is repeated 100 times with a different seed each time. When APs are deployed in a grid, channels are allocated to the APs in a way the co-channel interference is minimized. Table \ref{simPar} summarizes the simulation parameters.

\setlength\tabcolsep{10 pt}
	\begin{table}[ht]\centering

	\begin{small}
    \begin{tabular}{|c|c|c|c|}
  		\hline  
		\textbf{Parameter}& \textbf{Options}\\ \hline
	   \textbf{STA placement} & Uniform, Clustered\\ \hline
	   \textbf{AP placement} & Random, Grid\\\hline
	   \textbf{Throughput} & Fixed, Random\\\hline
	   \textbf{Channel bandwidth } & 20, 40, 80 MHz \\\hline
	   \textbf{$\mathbf{N_c}$} & 10 \\\hline
	   \textbf{Simulation seeds} & 100 \\\hline
	\end{tabular}	\caption{Scenario parameters.}
	\label{simPar}
	\end{small}
	\end{table}

\subsection{AP and STA placement effect}

In this section we will investigate the effect of STA and AP placement on the behavior of the \EG and \ES algorithms. We intend to find which scenarios can benefit the most when implementing either of the algorithms, as well as to confirm the viability of \EG and \ES with different number of STAs.

\subsubsection{Deployment distribution}

We start by studying the different AP and STA distributions and their effects on the performance achieved. We will study a scenario that consists of 16 APs and 64 STAs. The APs are placed at random or in a grid for uniform 
coverage, as most enterprise environments would do. The STAs are placed at random across the entire area or in clusters. Figure \ref{scenRand} shows a completely random deployment that follows a uniform distribution for both APs and STAs. Figure \ref{scenClust} shows a scenario with the 16 APs in a grid and the 64 STAs clustered in groups of 10 STAs. The second scenario is closer to a real environment, as users tend to group together in certain areas like cafeterias and meeting rooms. Each STA requests 4 Mbps and uses a 20 MHz channel. Both \EG and \ES use the same exploration value of $\varepsilon = 0.1$.

\begin{figure*}[ht]
\centering
   \begin{subfigure}[b]{.4\textwidth}
        \includegraphics[width=\textwidth]{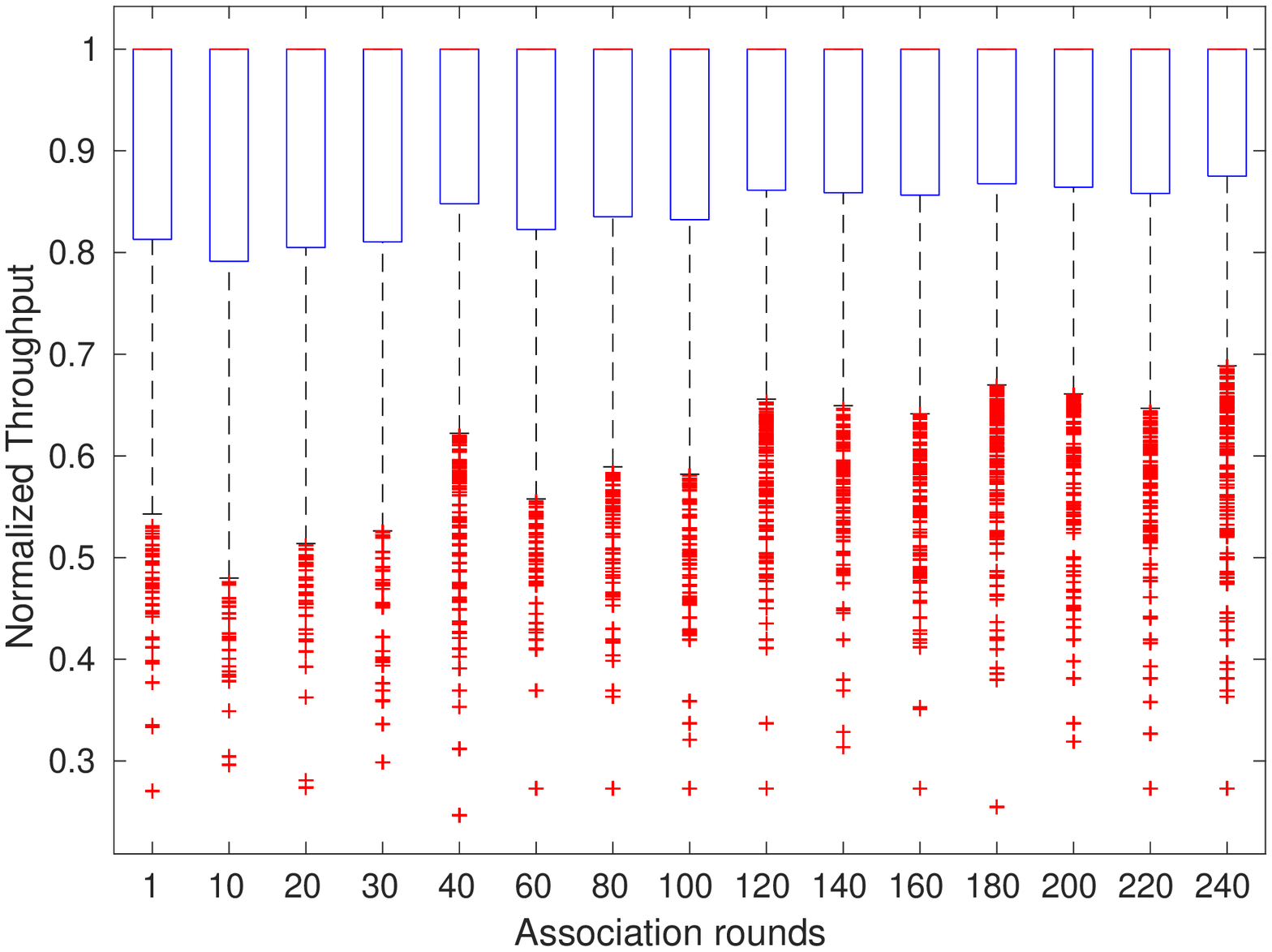}
    \caption{Evolution of \EG over time.}
    \label{streg}
    \end{subfigure}
    \begin{subfigure}[b]{.4\textwidth}
        \includegraphics[width=\textwidth]{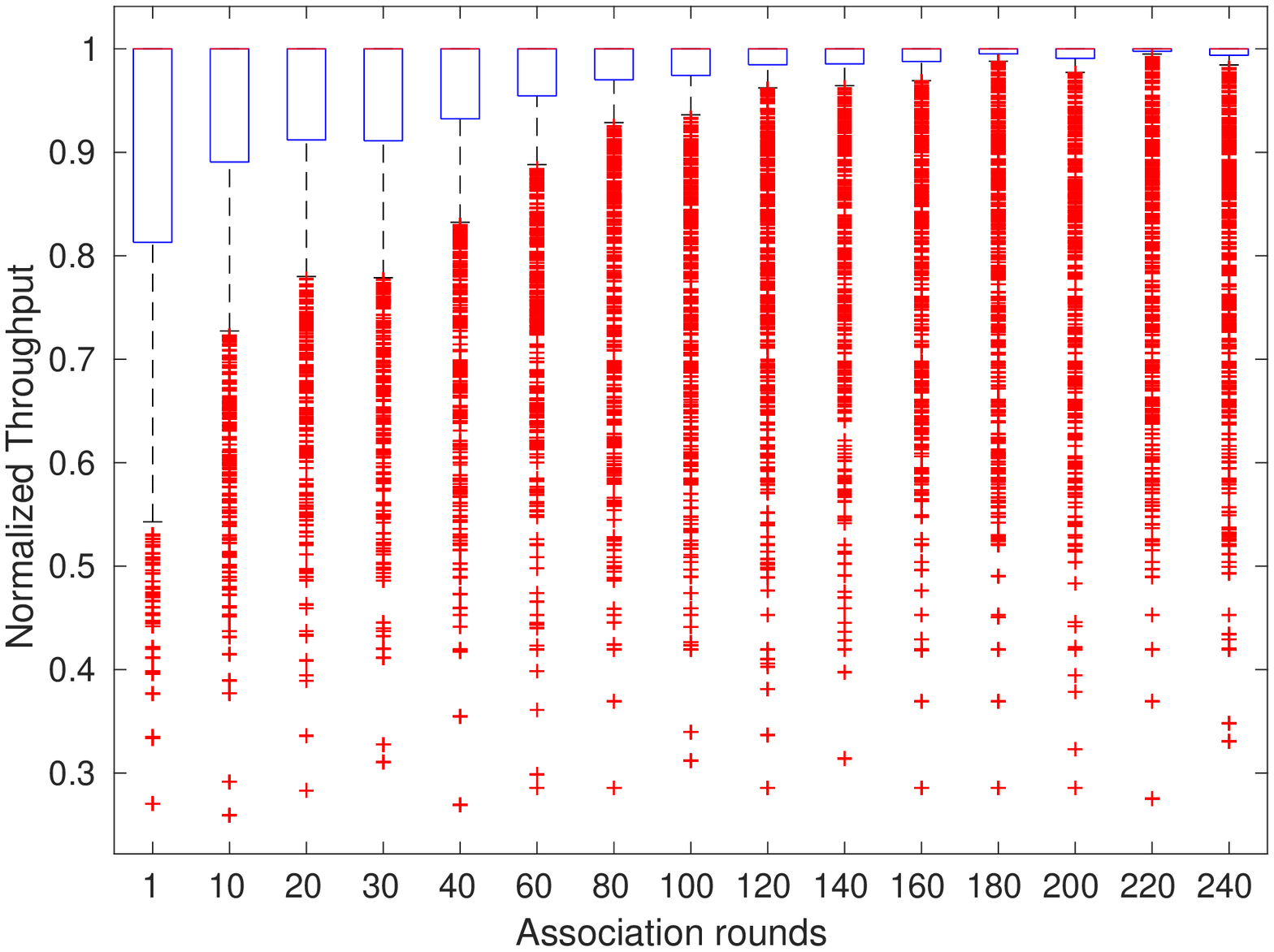}
    \caption{Evolution of \ES over time.}
    \label{stres}
    \end{subfigure}
     \begin{subfigure}[b]{.4\textwidth}
        \includegraphics[width=\textwidth]{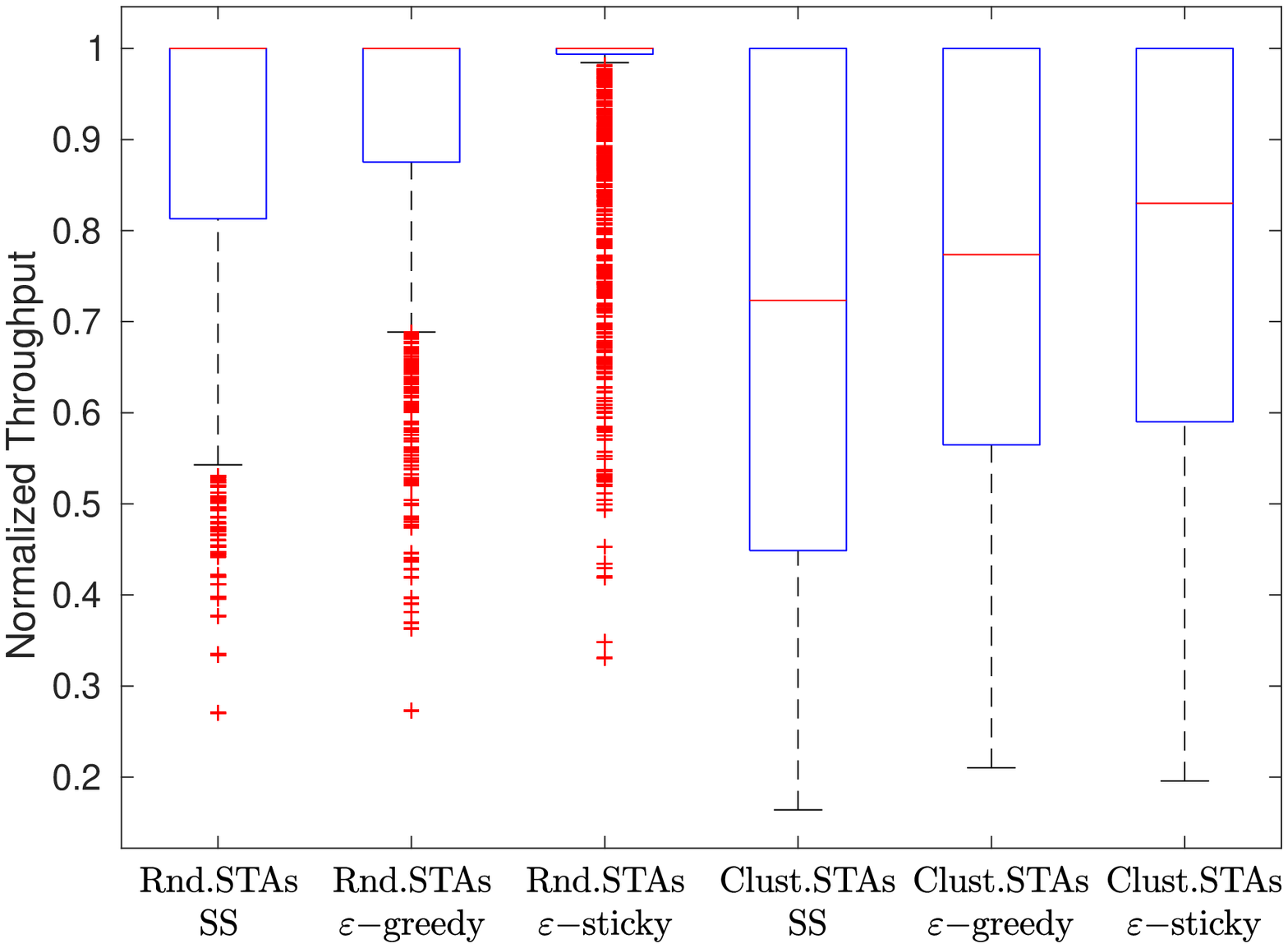}
    \caption{APs are distributed randomly.}
    \label{boxrandaps}
    \end{subfigure}
    \begin{subfigure}[b]{.4\textwidth}
        \includegraphics[width=\textwidth]{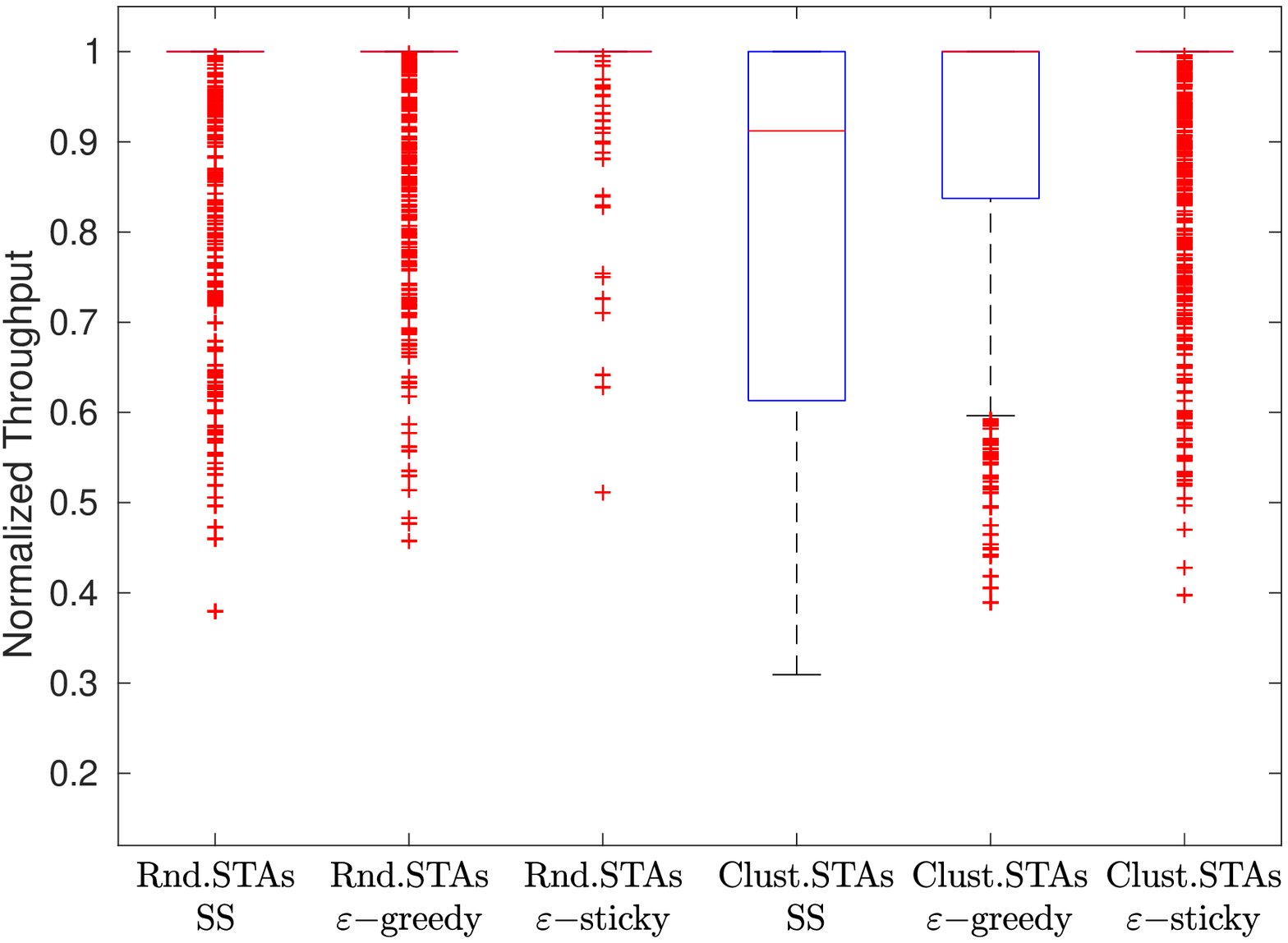}
    \caption{APs are in a grid.}
    \label{boxgridaps}
    \end{subfigure}
     \caption{Boxplots showing the distribution of the throughput for different scenarios.}
     \label{Fig:Scenariosss}
\end{figure*}

In Figure \ref{randbox} we deploy the APs randomly and compare the Strongest Signal (SS) method to both \EG and $\varepsilon$-sticky in both STA deployment cases. We can already observe with the SS method that when the STAs are uniformly distributed they achieve almost 90\% of their desired throughput, while distributing the STAs in clusters creates overcrowding issues on APs and leads to only achieving 70\% on average.
 
The next thing we can observe is that \EG improves upon the SS method in both cases. For the random STA distribution the throughput increases from 0.8899 to 0.9087 (a 2.1\% increase), and for the distribution in clusters it rises from 0.6948 to 0751 (8.08\% increase). Using \ES we can reach a higher throughput than with $\varepsilon$-greedy, reaching 0.9485 for random STA placement (6.58\% increase over SS) and 0.7777 for the clusters (increasing 11.93\% over SS).
 
We perform the same experiments with the APs distributed in a grid. Figure \ref{gridbox} shows \EG outperforming SS by 1.95\% and 12.65\% for the uniform and cluster deployment respectively, and \ES outperforming both of them, achieving an increase of 4.40\% and 17.96\% over Strongest Signal.  In both cases we can observe similar tendencies, with the deployment in clusters leading to a lower throughput than the random one. Further, when using clusters higher gains can be achieved. 

To better showcase the difference in the performance of both \EG and $\varepsilon$-sticky, we show the boxplots for both of them over several association rounds in Figures \ref{streg} and \ref{stres}. In them we can observe the range of throughput obtained by the STAs in a given association round, and how they both increase the 25th percentile over time, decreasing the overall variance. For \EG we can observe that while the median is stable, the 25th quartile and the minimum value are in constant fluctuation. This is a result of the exploration being constant over time, thus the STAs always have a chance of exploring and finding suboptimal association options. For \ES  the range of throughput values always decreases, until most STAs are satisfied in the last round. This is a result of the stickiness counter: as STAs explore, they may be able to find an AP that can serve their requested throughput, and once they have found it they stay on that AP as long as their throughput remains the same. As more STAs stick to their APs, the network becomes more stable due to less STAs exploring and changing APs, which allows the remaining STAs to better learn the state of the network and finding a feasible AP.

\begin{figure*}[ht]
\centering
   \begin{subfigure}[b]{.32\textwidth}
        \includegraphics[width=\textwidth]{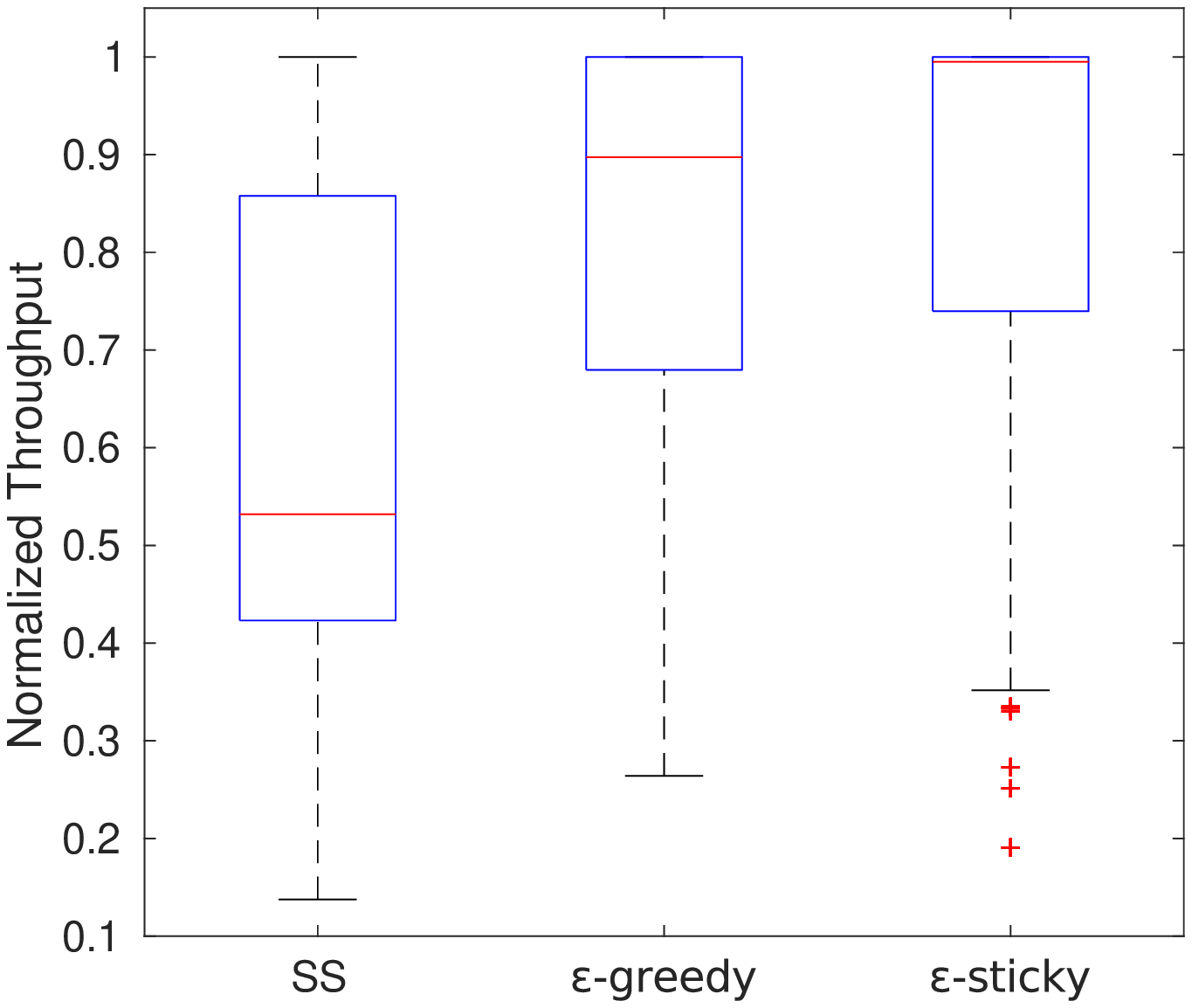}
    \caption{32 STAs, 8 Mbps.}
    \label{comp32}
    \end{subfigure}
     \begin{subfigure}[b]{.32\textwidth}
        \includegraphics[width=\textwidth]{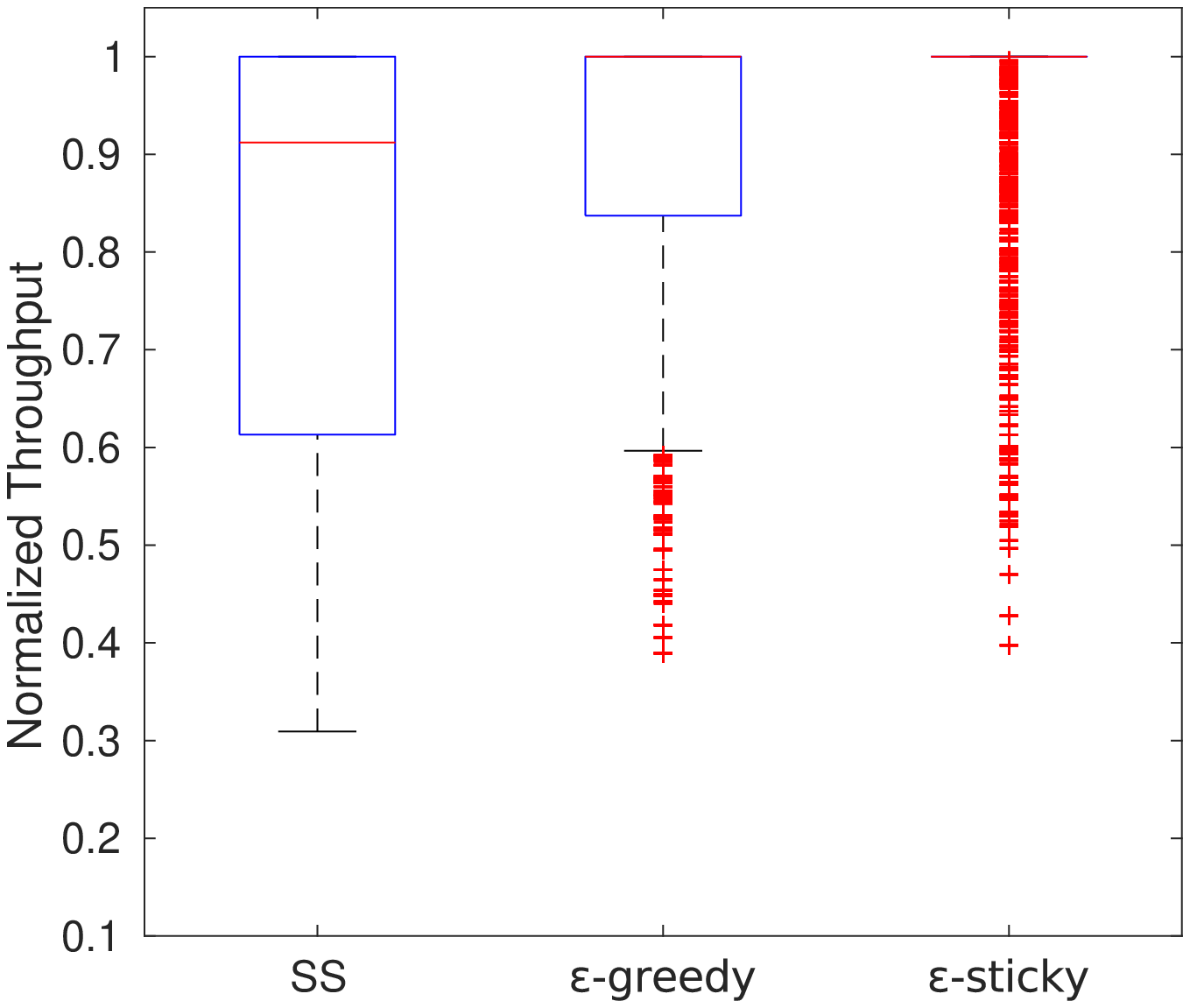}
    \caption{64 STAs, 4 Mbps.}
    \label{comp64}
    \end{subfigure}
    \begin{subfigure}[b]{.32\textwidth}
        \includegraphics[width=\textwidth]{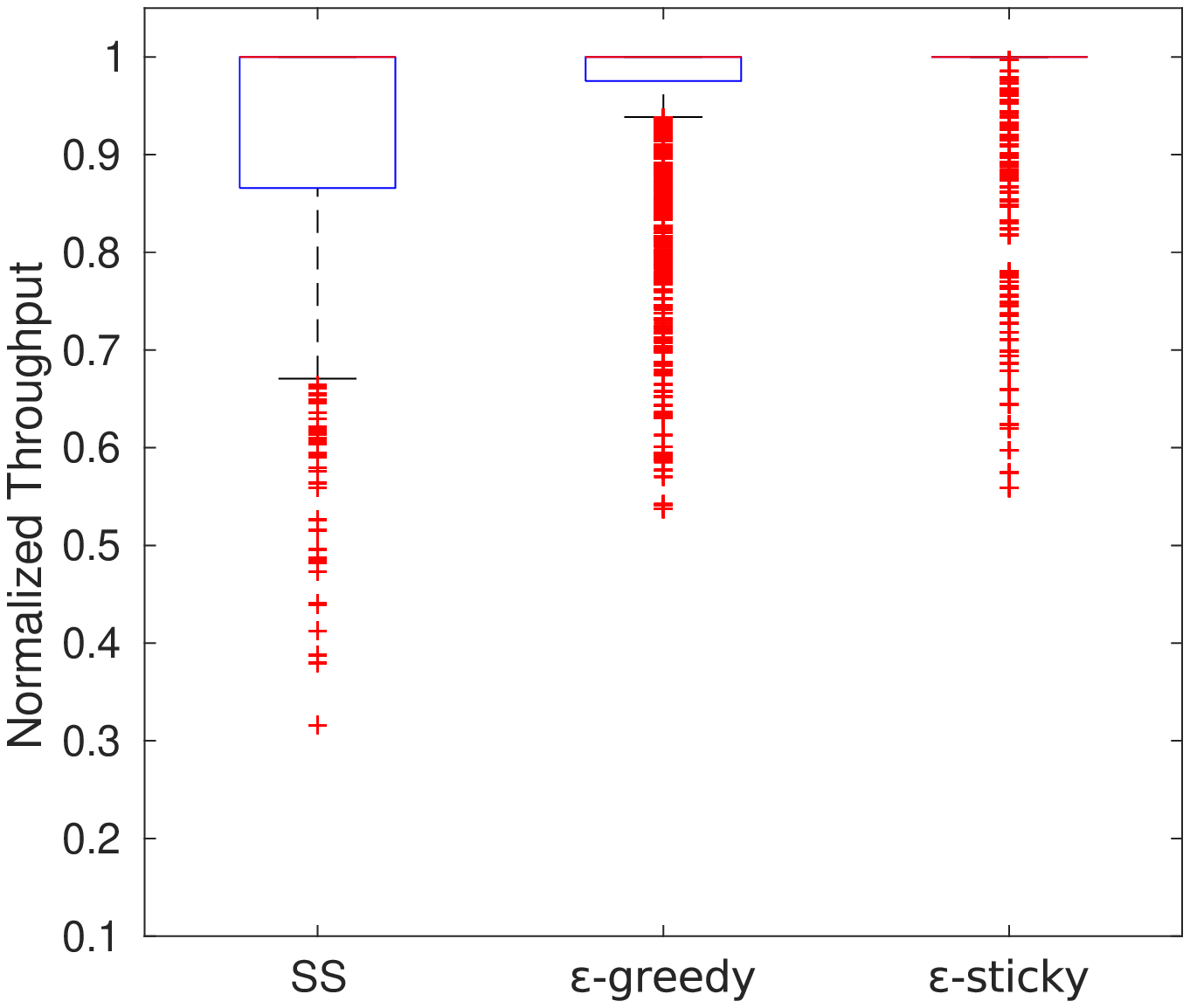}
    \caption{128 STAs, 2 Mbps.}
    \label{comp128}
    \end{subfigure}
     \caption{Normalized throughput when the number of STAs increase for APs in a grid and STAs in clusters. Note that the aggregate load is kept constant to 256 Mbps.}
     \label{Fig:Scenarios2}
\end{figure*}

Figures \ref{boxrandaps} and \ref{boxgridaps} show the boxplots for the last association round (round 240) of the simulation for each AP and STA distribution using each association method (note that the second and third boxplot of Figure \ref{boxrandaps} are the same as the last round for Figures \ref{streg} and \ref{stres}). We can observe that having the APs in a grid offers better performance as the ranges of normalized throughput values for all cases are higher and more compact in Figure \ref{boxgridaps}. Further, Figure \ref{boxgridaps} shows that with \ES we can reach throughput values for STAs in clusters that are very similar to those of distributing the STAs uniformly, greatly increasing the satisfaction of all STAs and only leaving a few outliers below 1.

One of the advantages of \ES over \EG is that the stickiness reduces the amount of times that a STA reassociates. Reassociations can be costly in terms of time and throughput, as all data flows are stopped during the process of switching from one AP to another, until the new AP authorizes the STA to access the network. With \ES we can find a solution for the STA while not interrupting the service quite as often as for $\varepsilon$-greedy. For the scenarios considered, we show the reassociation ratio between \EG and \ES in Table \ref{reassoc} (i.e., the number of reassociations caused by \EG divided by the ones caused by $\varepsilon$-sticky). In the worst case of random APs and STAs in clusters we find that there are almost twice the reassociations with \EG than with $\varepsilon$-sticky, and in the case of APs in a grid and STAs in clusters we find an impressive ratio of 64 \EG reassociations for a single \ES one.

\setlength\tabcolsep{5 pt}
	\begin{table}[ht]\centering
	\begin{small}
    \begin{tabular}{|c|c|c|c|c|}
  		\hline  
		\textbf{AP dist.}& \textbf{STA dist.}& \textbf{$\varepsilon$-g Reassoc./ $\varepsilon$-s Reassoc.}\\ \hline
	   Random & Random & 6.64 \\\hline
	   Random & Clusters & 1.79 \\\hline
	   Grid & Random & 35.23 \\\hline
	   Grid & Clusters & 64.84 \\\hline
	\end{tabular}
	\caption{Ratio between the number of reassociations carried out by \EG versus $\varepsilon$-sticky. }
	\label{reassoc}
	\end{small}
	\end{table}

From these results we can infer that having the APs in a grid allows for higher throughput gains, as the grid formation creates fewer instances of STAs with a bad signal by covering the whole area evenly. Regarding the STA distribution, we find that distributing them uniformly at random over the entire area also spreads them evenly among all APs, thus resulting in a low number of unsatisfied users. When the STAs are distributed in clusters, the network has a lower average throughput, and it is in those cases where our algorithms can obtain the highest improvement over the default association method. As we have seen in this section, deploying the APs in a grid and having a chaotic distribution of STAs creates more opportunities for both \EG and \ES to find a better association scheme. Finally, it bears mentioning that both algorithms achieve better results than SS in very few iterations, and by the 20th association round (1 hour) we can see a very clear improvement.

\subsubsection{Number of STAs in the network}

We continue by studying the effect of changing the number of STAs in the network. We will now test the same 16 AP configuration but will deploy 32, 64 and 128 STAs. We modify the user requirements to keep the network load the same so that we can focus on the effect of the number of STAs deployed. For each of these 3 scenarios, every single user requests 8, 4  and 2 Mbps, respectively. All simulations deploy the APs in a grid and the STAs in clusters, as we have shown before that those scenarios are the ones with a greater potential for improvement.

The results can be found in Figure \ref{Fig:Scenarios2}, where we show the last association round for each case.  We can observe that \EG always improves upon the SS method and that \ES consistently outperforms them both.

For the case of 32 STAs in Figure \ref{comp32} we can observe that the entire range of throughput values for both \EG and \ES is higher than the one for SS, with both methods having the 75th percentile reaching 1 and their medians being higher than the 75th percentile of SS. For 64 STAs in Figure \ref{comp64} we can observe that both \EG and \ES reach a median of 1 while the one for SS is  91.12\%. The 25th percentile is also greatly improved, going from 61.13\% to 83.72\% and 1 for \EG and \ES respectively. Figure \ref{comp128} shows the scenario with 128 STAs, in which most STAs are already satisfied with a median of 1 with SS and a high 25th percentile of 86.5\%. But even in this case we can increase this percentile to 97.53\% for \EG and 1 for $\varepsilon$-sticky. 

 In these particular cases, the reassociation ratio between \EG and \ES is 2.88 for the 32 STAs, 64.84 for 64 STAs and 608.62 for 128 STAs, which shows the much higher efficiency of \ES in terms of achieving higher throughput with a minimal number of reassociations.

\begin{figure*}[ht]
\centering
   \begin{subfigure}[b]{.329\textwidth}
        \includegraphics[width=\textwidth]{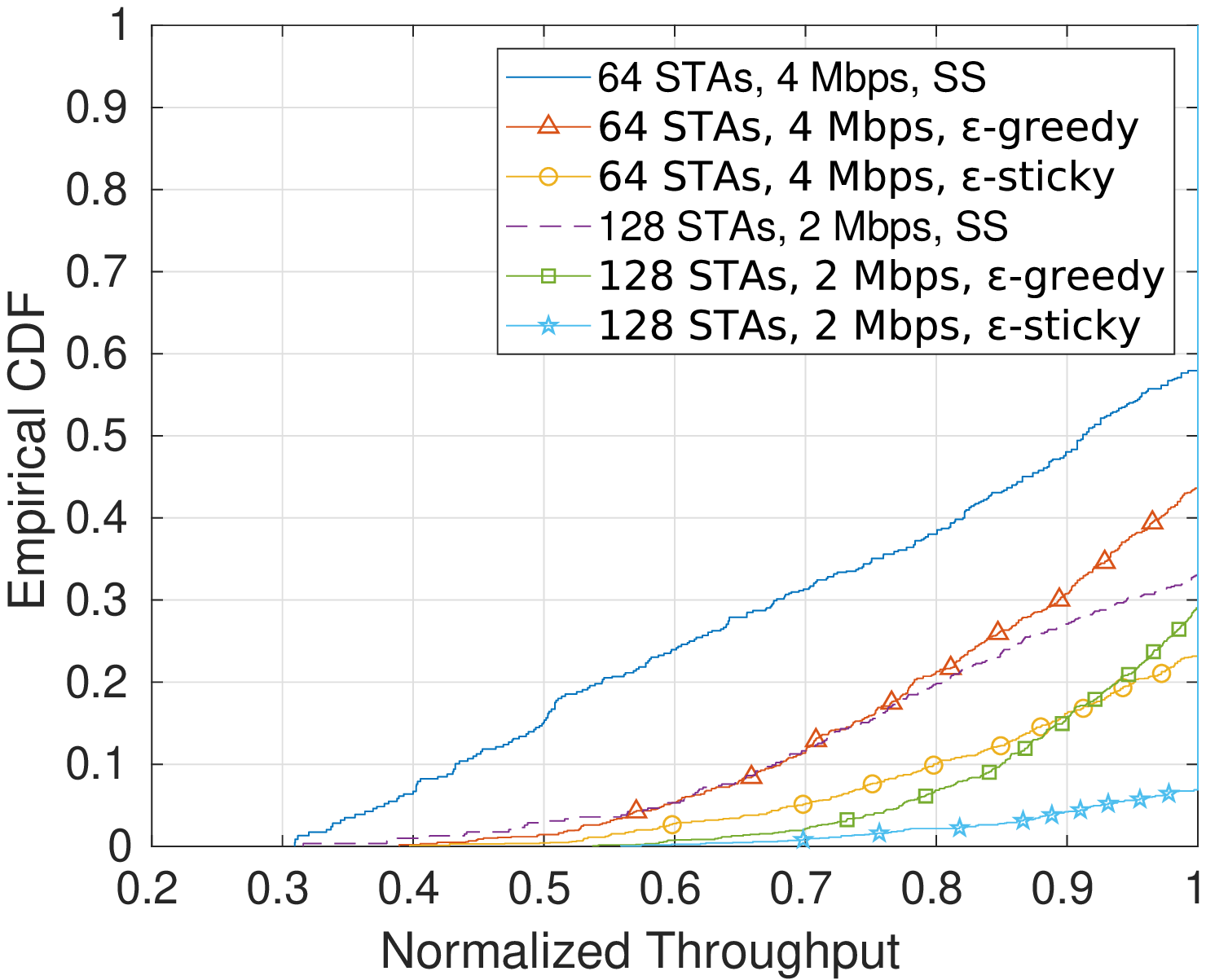}
    \caption{20 MHz. }
    \label{20mhz}
    \end{subfigure}
     \begin{subfigure}[b]{.329\textwidth}
        \includegraphics[width=\textwidth]{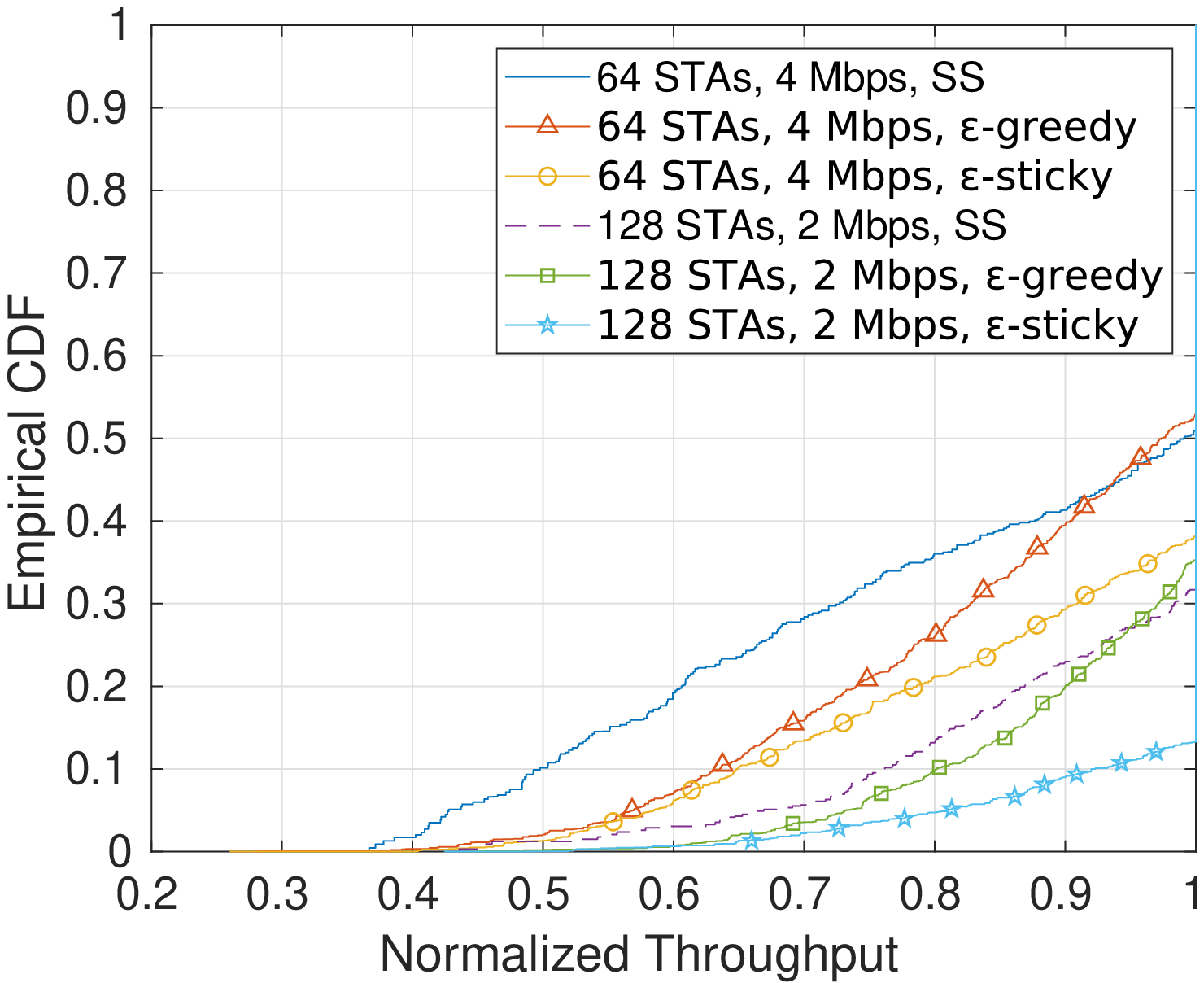}
    \caption{40 MHz.}
    \label{40mhz}
    \end{subfigure}
    \begin{subfigure}[b]{.329\textwidth}
        \includegraphics[width=\textwidth]{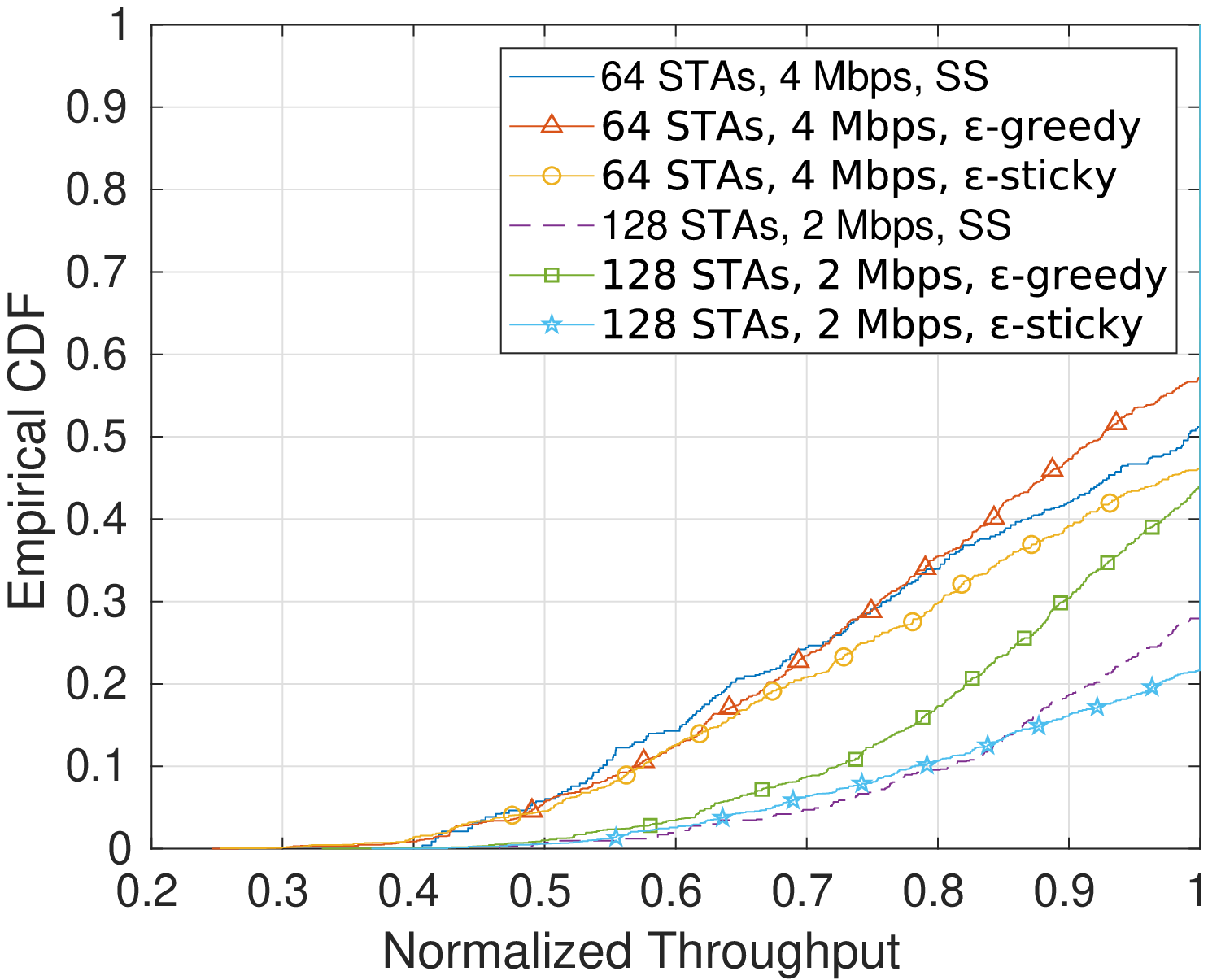}
    \caption{80 MHz.}
    \label{80mhz}
    \end{subfigure}
     \caption{Empirical CDFs for 20, 40 and 80 MHz channels.}
     \label{F1}
\end{figure*}

While the network load was kept at the same amount, we can observe that the ranges of throughput are different in these three cases. The 32 STAs scenario with 8 Mbps of load per user shows lower throughput values, while the 128 STAs scenario and 2 Mbps per user shows higher ones. This is a consequence of the airtime required for the load requested by the STAs. Since a STA requesting 2 Mbps requires less airtime than one requesting 8 Mbps. The scenario for 128 STAs gives more opportunities to find a feasible solution, as changing a STA from an AP to another will have a smaller impact than those of the 32 STA scenario. Even considering this, both \EG and \ES achieve successful results in all scenarios, showing that the number of STAs in a network is never a detriment to the performance of both algorithms.

\subsection{Channel bonding}

In this section we aim to ascertain the effects of using a different amount of channels, as well as the impact that using a different channel bandwidth can have on our algorithms. The bandwidths and channels used are specified in Table \ref{chann}. We use the same amount of spectrum in all cases, meaning that as we increase the bandwidth we have less orthogonal channels available for the APs. We sill use the same simulation parameters as before, with 16 APs in a grid and the STAs distributed in clusters. We will consider two scenarios, with 64 and 128 STAs.

\setlength\tabcolsep{5 pt}
	\begin{table}[ht]\centering
	\begin{small}
    \begin{tabular}{|c|c|c|c|c|}
  		\hline  
		\textbf{Channel bandwidth}& \textbf{Channels used}\\ \hline
	   20 MHz & 36, 40, 44, 48, 52, 56, 60, 64\\\hline
	   40 MHz & 38, 46, 54, 62 \\\hline
	   80 MHz & 42, 58 \\\hline
	\end{tabular}
	\caption{Channels used for each bandwidth.}
	\label{chann}
	\end{small}
	\end{table}

Figure \ref{F1} shows the empirical Cumulative Distribution Functions (CDFs) of the last association round for the three association methods with the different channel distributions. Figure \ref{20mhz} shows the cdfs when using 20 MHz channels. For 64 STAs all CDFs are quite different, with SS showing 57.93\% of the STAs receiving less than the throughput requested, and a minimum normalized throughput value of 0.309. With \EG we find that 43.62\% of the STAs do not reach their desired throughput, while the minimum throughput achieved is 0.3831. With \ES the minimum throughput is further increased to 0.3974 and now only 23.17\% of all STAs are not satisfied. In this particular scenario we can observe that all STAs receive higher throughput with \EG and $\varepsilon$-sticky. 

In the case of 128 STAs we find similar results. The minimum throughput achieved goes from 0.3158 with SS, to 0.5373 with \EG and 0.5588 with $\varepsilon$-sticky. And the amount of unsatisfied STAs for each method is 33.12\%, 29.05\% and 6.88\% respectively. In this case, much like before, \EG improves upon SS and \ES improves upon $\varepsilon$-greedy. 

If we look at the CDFs for 40 MHz channels in Figure \ref{40mhz} we find some strange occurrences when comparing SS and $\varepsilon$-greedy. While for the most part STAs using \EG obtain higher throughputs than those using SS, we can observe that for both 64 and 128 STAs, the \EG CDF increases faster than that of the SS method, finally overtaking it and leading to the \EG method having more STAs that do not reach their requested throughput, 52.93\% for \EG and 50.9\% for SS in the 64 STA scenario, and 35.26\% and 31.68\% in the 128 STA scenario. For \ES we find that the throughput is always higher than with the other methods, and that there are more STAs fully satisfied as well.

Figure \ref{80mhz} shows the CDFs for channels of 80 MHz. This Figure follows the tendency of the prior one, with the CDFs for \EG and \ES being closer to those of SS, and in fact we can observe that the performance of \EG for the 128 STA scenario is low, leaving 43.94\% of STAs without their desired throughput, while this only happens to 27.93\% of STAs with SS. For both scenarios, \ES manages to outperform SS, but in the case of 64 STAs the three CDFs are very similar.

Channel reuse has a huge impact on the performance of \EG and $\varepsilon$-sticky. As the amount of channels available decreases, the amount of APs and STAs sharing the same channel increases, leading to situations in which a STA may have multiple APs from which to choose but no alternatives in terms of channels. A STA may reassociate to another AP, but if it is sharing the channel with the previous AP, that choice is meaningless as there is nothing to gain from that AP change. Using higher bandwidth channels allows for higher data rates, but we have shown that these higher data rates cannot compensate for the loss of orthogonal channels. For our algorithms to work optimally, it is imperative that most of the APs in range of a STA use different channels, or the reassociation will only lead to a lower data rate and a worsening of the network.

\subsection{Variable loads}

In previous sections, we have considered that the load of the STAs was a fixed value for all association rounds, and that all STAs used the same value. In this section we will allow the STAs to select their load at each association round from a range of load values, thus creating a more chaotic environment. The goal is to observe if the proposed framework is able to keep up its performance with the added randomness.

\begin{figure}[h]
    \centering
    \includegraphics[width=0.48\textwidth]{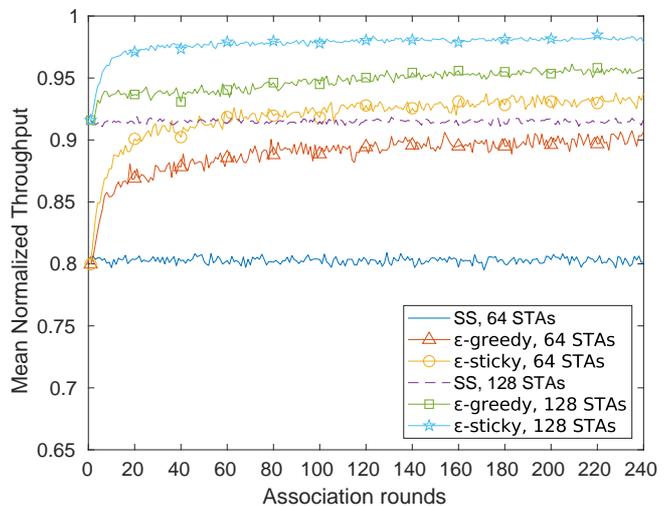}
    \caption{Mean normalized throughput obtained for the variable load case.}
    \label{F5}
\end{figure}

\begin{figure*}[ht]
\centering
   \begin{subfigure}[b]{.435\textwidth}
        \includegraphics[width=\textwidth]{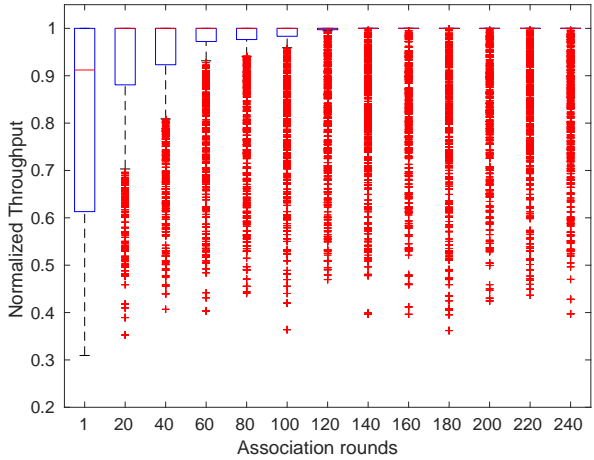}
    \caption{Fixed load.}
    \label{thr20}
    \end{subfigure}
     \begin{subfigure}[b]{.44\textwidth}
        \includegraphics[width=\textwidth]{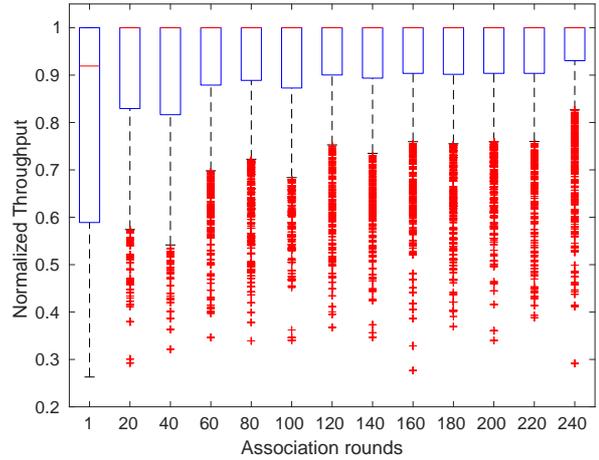}
    \caption{Variable load.}
    \label{thr40}
    \end{subfigure}
     \caption{Fixed vs variable loads for 16 APs in a grid and 128 STAs in clusters.}
     \label{F2}
\end{figure*}

For the next simulations we will replicate the previous scenarios of 64 and 128 STAs. Instead of having a fixed load of 4 and 2 Mbps per STA, they will now have an average load of 4 and 2 Mbps respectively. To do this each STA will be allowed to take values from the range of $[1, (\text{Avg. Load}\cdot{2})-1]$ Mbps. For each association round, each STA will choose a new value from this range uniformly at random.

Figure \ref{F5} shows the throughput for each scenario and association method. We can observe the added randomness is present in the SS method, where there are now  fluctuations that were not present in the previous tests. These fluctuation are present also in all other algorithms. We can observe that this deviation is lower when using \ES than when using $\varepsilon$-greedy, thanks to the sticky counter.

For the scenario with 64 STAs we can compare Figure \ref{F5} directly to part of Figure \ref{gridbox}, where we had 64 STAs in clusters but with a constant 4 Mbps request. The results do not seem to be heavily affected by the random load, as \EG can achieve a 13.15\% throughput improvement over SS (whereas we achieved 12.65\% with static load) by the last association round, while \ES reaches a 16.98\% increase (similar to the 17.96\% obtained before). If we deploy 128 STAs we find that \EG and \ES also perform as expected, with \EG obtaining a 4.4\% increase in throughput and \ES a 6.78\%.

\begin{figure*}[ht]
\centering
   \begin{subfigure}[b]{.44\textwidth}
        \includegraphics[width=\textwidth]{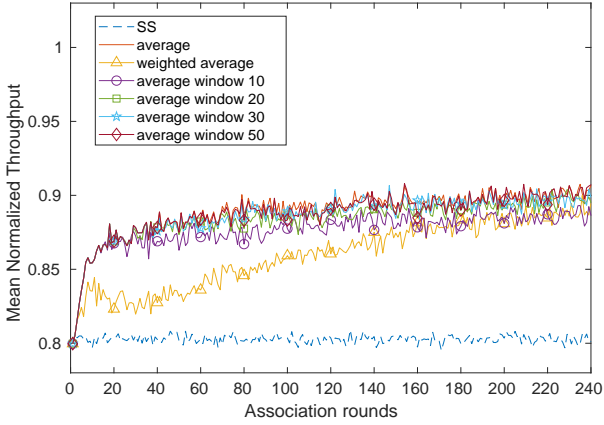}
    \caption{$\varepsilon$-greedy.}
    \label{egavgcom}
    \end{subfigure}
     \begin{subfigure}[b]{.44\textwidth}
        \includegraphics[width=\textwidth]{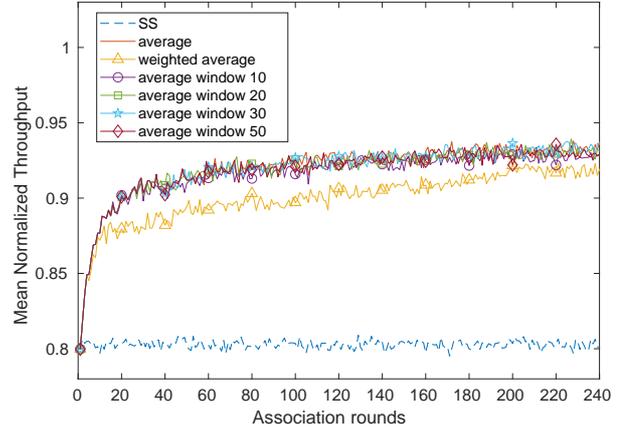}
    \caption{$\varepsilon$-sticky. }
    \label{esavgcom}
    \end{subfigure}
    
     \caption{Mean normalized throughput for different strategies to compute the reward for each AP.}
     \label{F12}
\end{figure*}

Figure \ref{F2} better showcases the differences between static and random load with the boxplot for \ES in the 128 STAs scenario. Figure \ref{thr20} shows the static load scenario, where we can observe that the range of values diminishes until all STAs outside of outliers achieve their requested throughput. Figure \ref{thr40} shows the random case, in which we find that \ES takes longer to reduce the throughput range, and ends up with a much wider range of values than the static case.

On average, it would seem that this added level of randomness has little effect on our simulations, but adding random load actually prevents most scenarios from having all STAs satisfied, as a configuration that satisfies all STAs may be valid on one association round but not on the next one. Even with this added difficulty, both \EG and \ES are capable of learning and outperform the standard association method.

\subsection{Reward computation}

\begin{figure*}[ht]
\centering
   \begin{subfigure}[b]{.44\textwidth}
        \includegraphics[width=\textwidth]{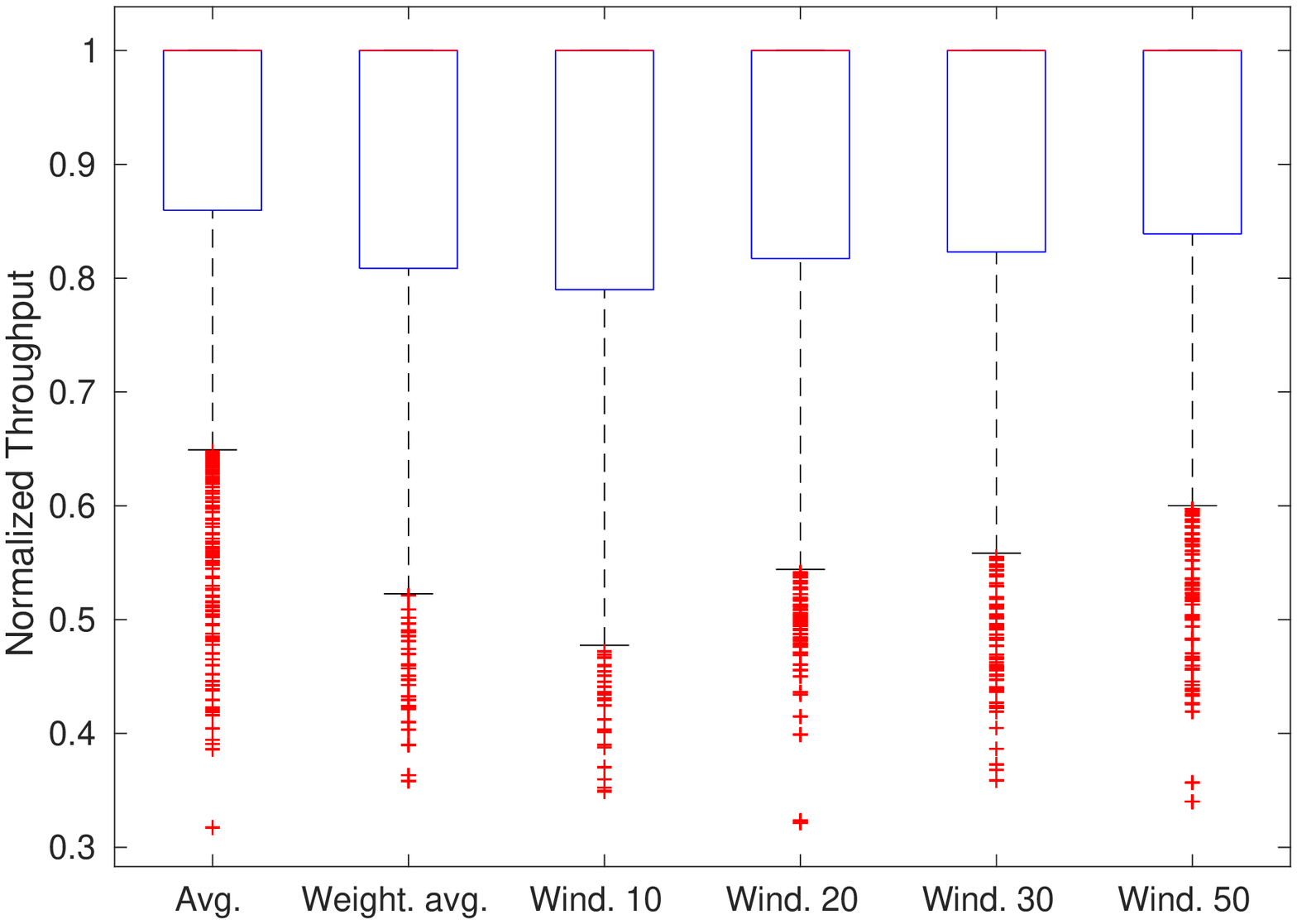}

    \caption{$\varepsilon$-greedy.}
    \label{egavgcomb}
    \end{subfigure}
     \begin{subfigure}[b]{.44\textwidth}
        \includegraphics[width=\textwidth]{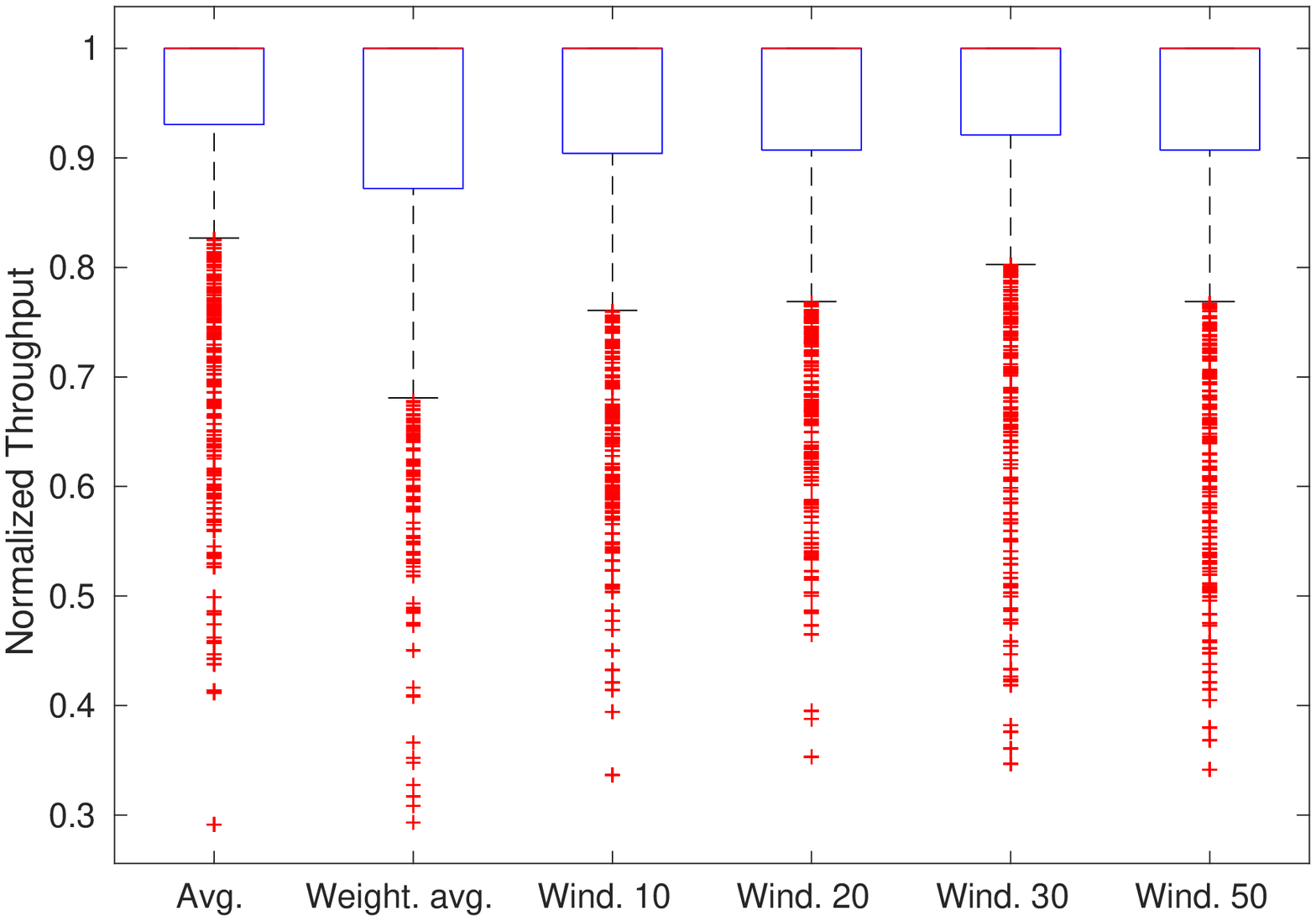}
    \caption{$\varepsilon$-sticky.}
    \label{esavgcomb}
    \end{subfigure}
    
     \caption{Boxplots of the reward in the last association round for different strategies.}
     \label{F14}
\end{figure*}
\begin{figure*}[ht]
\centering
   \begin{subfigure}[b]{.44\textwidth}
        \includegraphics[width=\textwidth]{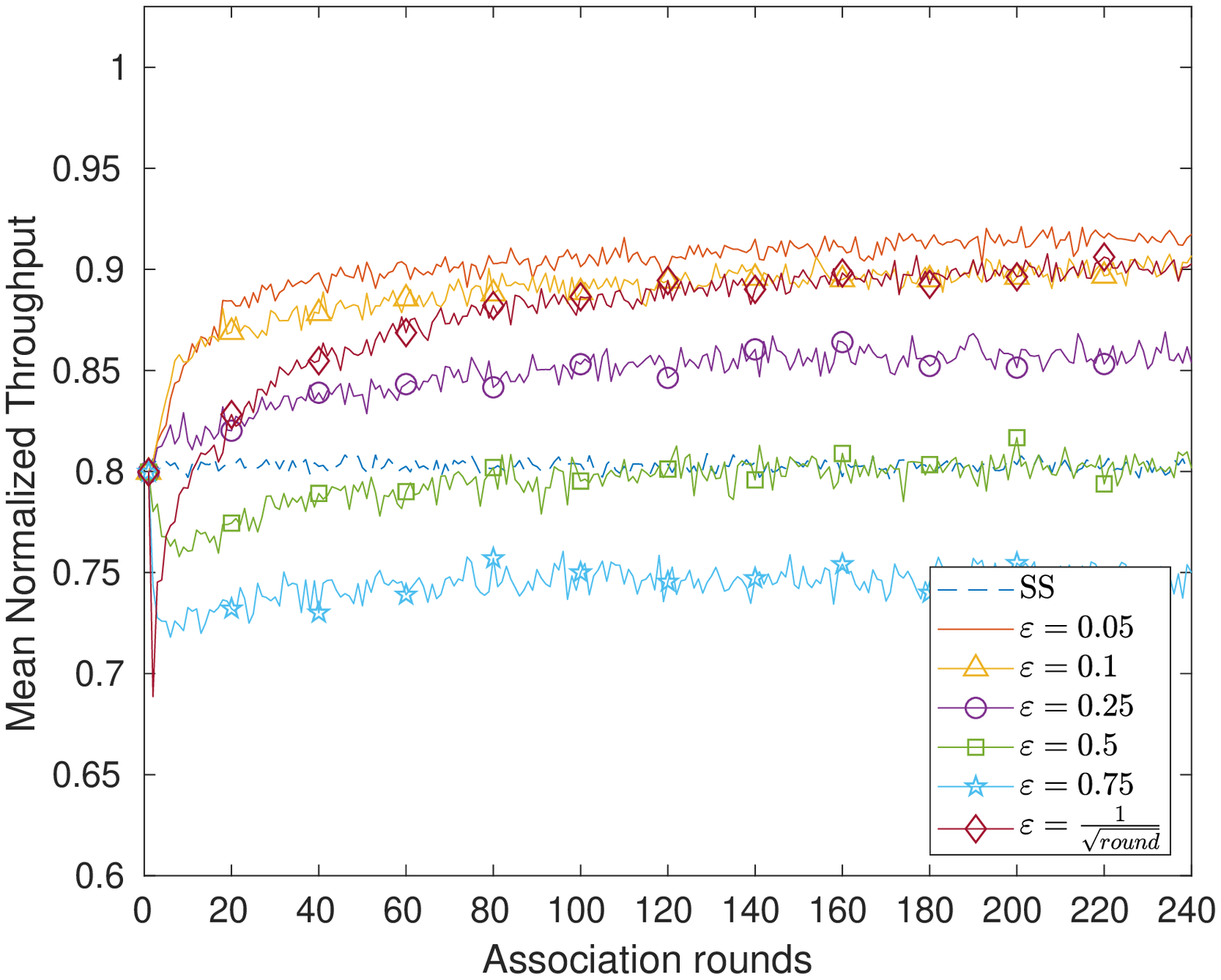}
    \caption{$\varepsilon$-greedy.}
    \label{epsilong}
    \end{subfigure}
     \begin{subfigure}[b]{.44\textwidth}
        \includegraphics[width=\textwidth]{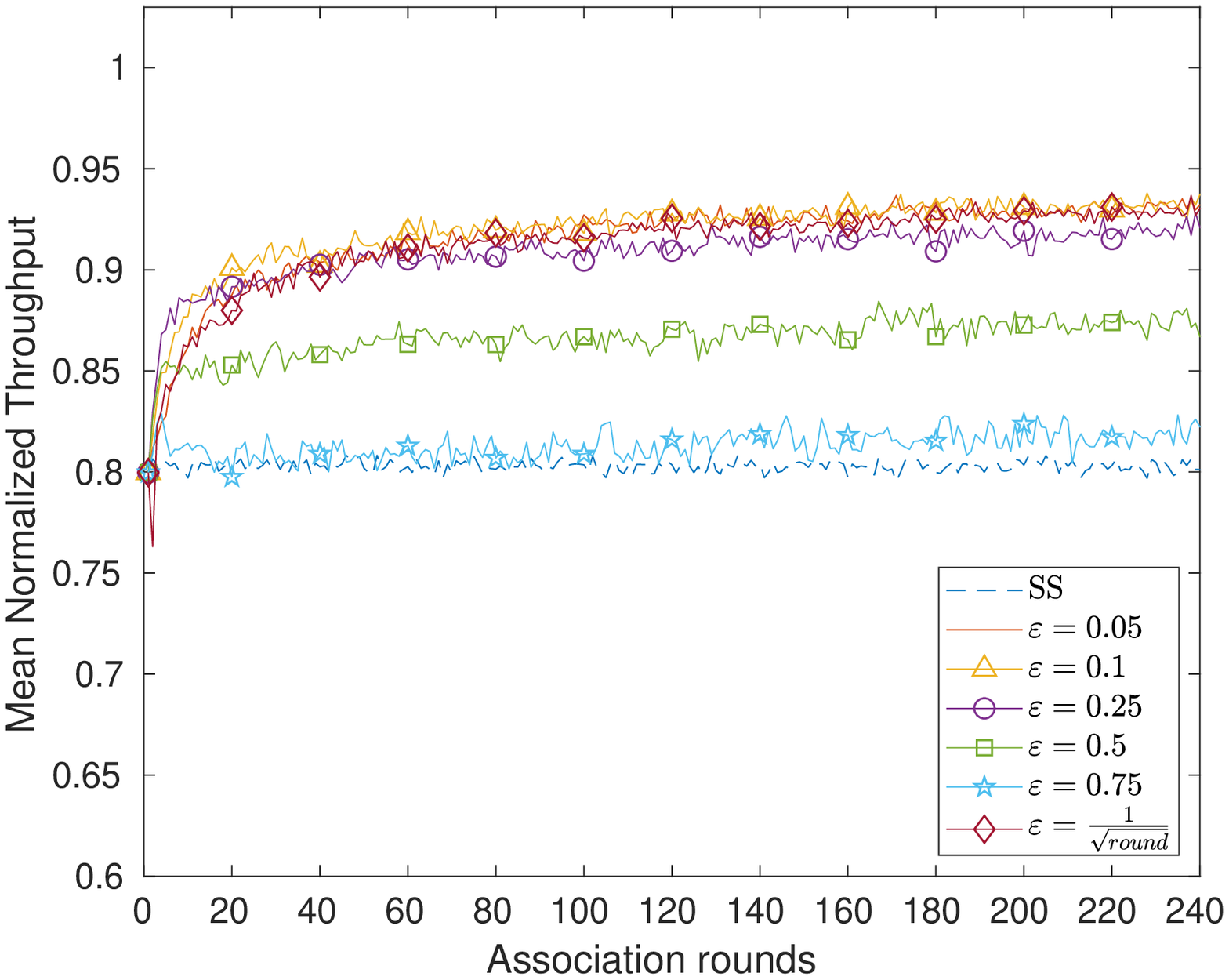}
    \caption{$\varepsilon$-sticky.}
    \label{epsilons}
    \end{subfigure}
     \caption{Effect of different values of $\varepsilon$ in \EG and \ES learning capabilities.}
     \label{Fe}
\end{figure*}
\begin{figure*}[htb]
\centering
   \begin{subfigure}[b]{.44\textwidth}
        \includegraphics[width=\textwidth]{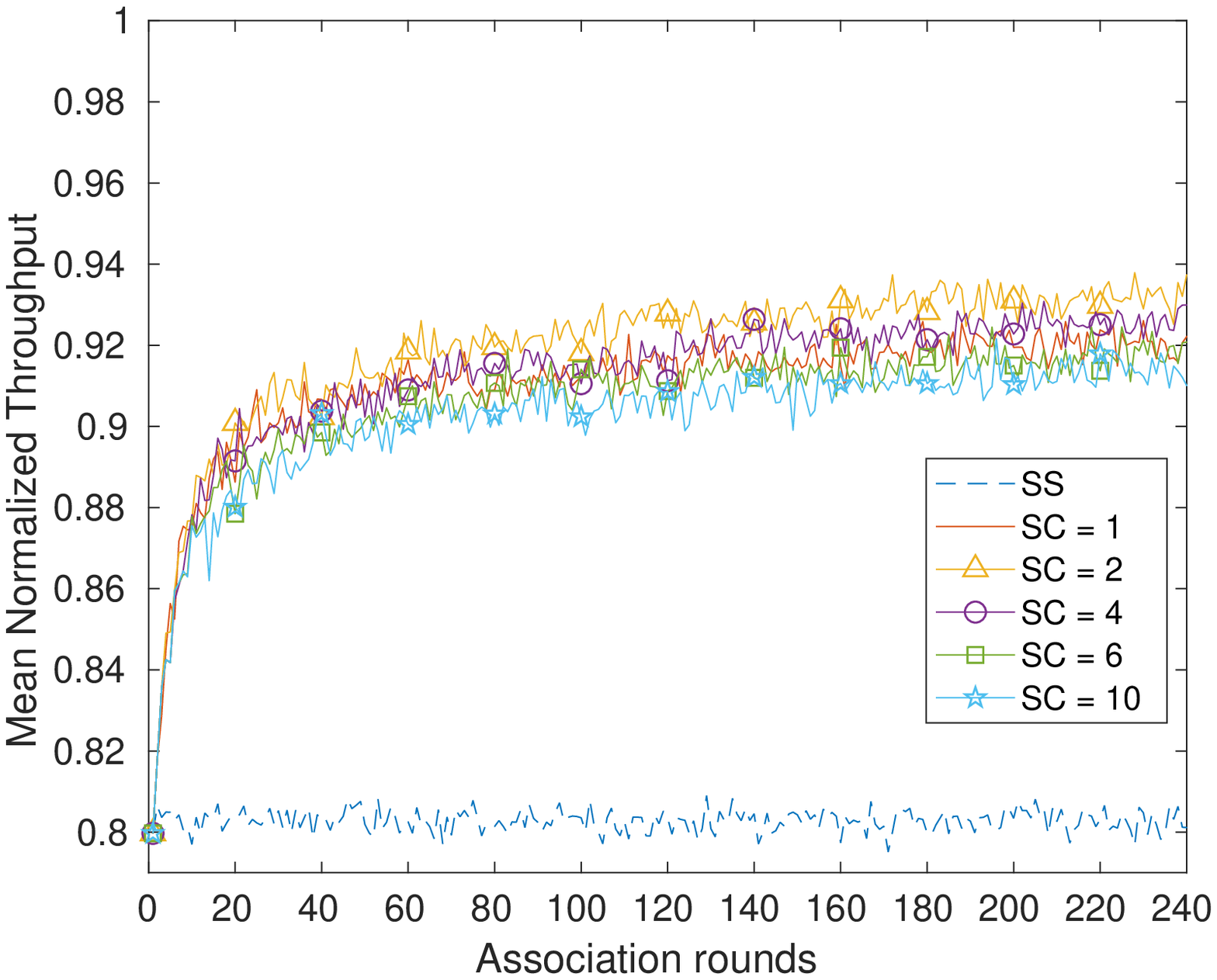}
    \caption{Temporal evolution of the mean normalized throughput for different sticky counter values.}
    \label{stickycoplot}
    \end{subfigure}
     \begin{subfigure}[b]{.44\textwidth}
        \includegraphics[width=\textwidth]{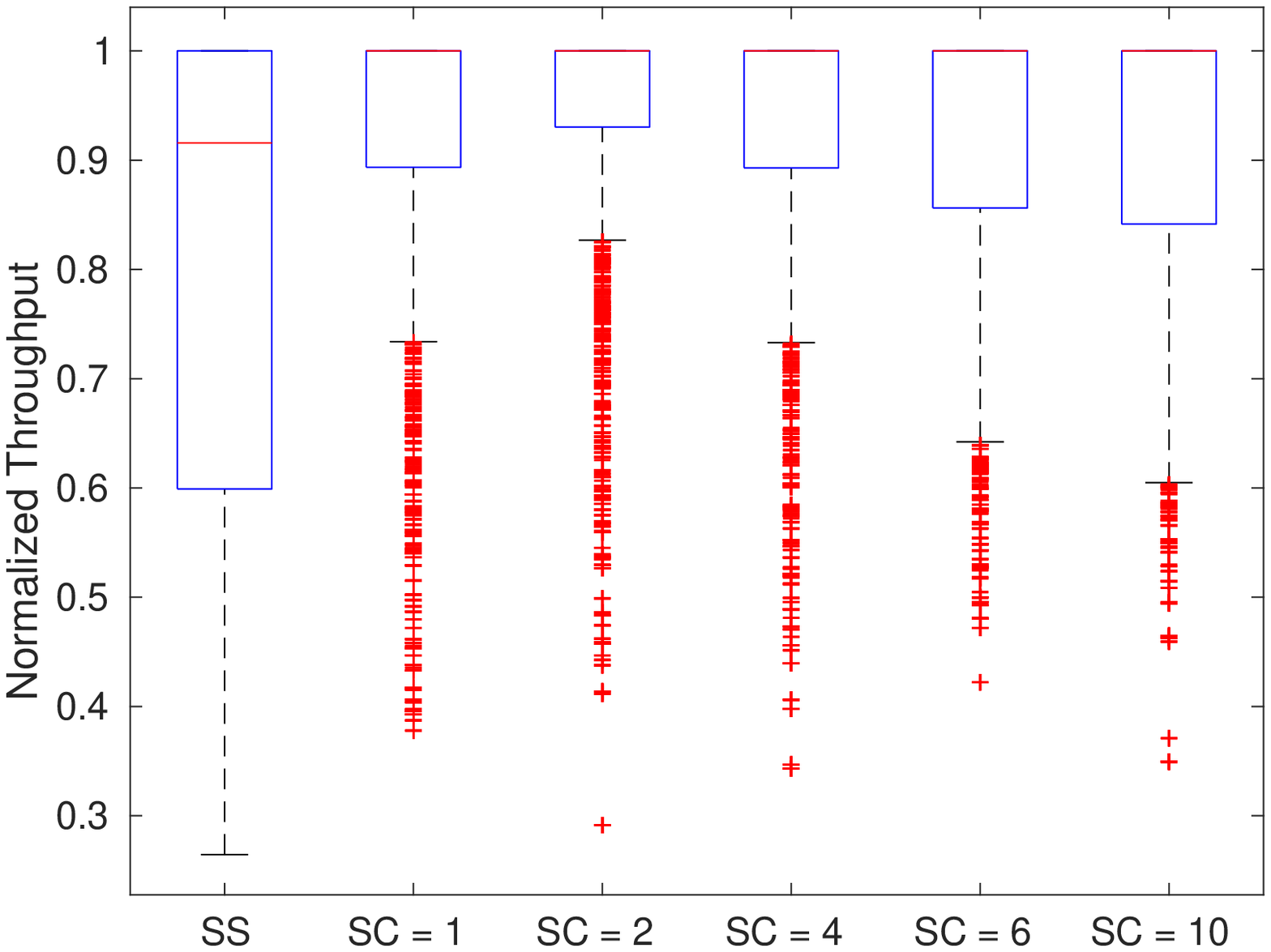}
    \caption{Boxplot of the normalized throughput of the last association round.}
    \label{stickycobox}
    \end{subfigure}
     \caption{Impact of the sticky counter value in the learning capabilities of $\varepsilon$-sticky.}
     \label{Fsc}
\end{figure*}
In this section we will discuss different ways to compute the reward used by the $\varepsilon$ algorithms. We will consider the following:

\begin{itemize}
    \item \textbf{Average:} All previous rewards of an AP are averaged according to the number of association rounds spent on said AP. 
    \item \textbf{Weighted:} Each reward is weighted prior to their addition. The most recent value gets a weight of $1$ and each subsequent value gets its weight by $1- \frac{x}{n}$, where $x$ is the reward iterator ($x = 1, 2,..$) and $n$ the number of rewards.
    \item \textbf{Use of a window:} An average in which we apply a limitation over the amount of rewards considered (i.e., we ignore rewards that are too old).
\end{itemize}

Once again, we place 16 APs in a grid and 64 STAs deployed in clusters, each requesting an average load of 4 Mbps. 

Figure \ref{F12} shows the results for each method and \EG and $\varepsilon$-sticky. While the tendencies are the same, we can better observe their performance in Figure \ref{egavgcom} for $\varepsilon$-greedy. There, we find that the weighted average performs the worst out of all of them, requiring more rounds than the other methods to reach the same level of throughput. The use of a window of 10 rounds is the second worst method, staying slightly below the performance of the windows of 20, 30, 50, or using every single reward, which all achieve very similar results. The results for \ES in Figure \ref{esavgcom} work in the same manner, but there the window of 10 rounds seems to work as well as the others.

We compare the boxplots of the last association round for each method in Figure \ref{F14}, where the differences between the reward methods are a bit clearer. For $\varepsilon$-greedy, in Figure \ref{egavgcomb}, we find that there is a difference in the 25th percentile and minimum values for all methods, with the regular average having the highest of all of them. We can also observe that in this round, the recency average outperforms the window of 10 rounds. It is also worth noting that the higher the window of the average, the higher the value of the 25th percentile and minimum.  

For $\varepsilon$-sticky, in Figure \ref{esavgcomb}, we find a different tendency. The weighted average is still the worst method. Also, we can observe that the window of 20 and 50 rounds have almost identical performance, with the window of 30 rounds slightly outperforming them both. The best performance, much as before, comes from the full average.

In this section we have considered weighting the average and limiting the amount of rewards used for the computation of the reward. While ultimately we found that using the average of all our rewards was the best option, we also found that using a window of 20 or 30 rounds can achieve very similar results. In this way, using these windows could be key in scenarios where the network conditions change abruptly (i.e., the number of STAs or traffic patterns change with the time).

\subsection{Parameter Optimization}

In this section we want to study the impact that the  value of $\varepsilon$ and the sticky counter have on the achievable throughput, and on the learning capabilities of \EG and $\varepsilon$-sticky. We use the 64 STAs scenario, with variable load of mean 4 Mbps. We perform 100 simulations for each value of $\varepsilon$ and SC.

Figure \ref{Fe} compares the throughput achieved for several values of $\varepsilon$, including the case where $\varepsilon$ decreases over time. For \EG in Figure \ref{epsilong} we find that a lower exploration rate leads to better results. The reason is that it reduces the amount of changes in the network every round, meaning that the information gathered by the STAs is more relevant. When $\varepsilon$ is high, the network becomes chaotic with a lot of STAs reassociating at each round, which in turn leads to the agents acquiring information that will not be useful in the next rounds. This effect is best observed in the decreasing $\varepsilon$, as the agents start with a high exploration rate and the throughput achieved initially goes below that of the SS method. As the exploration rate decreases, the agents can recover over time and achieve similar results to low fixed $\varepsilon$ values. 

Figure \ref{epsilons} shows the same comparison for $\varepsilon$-sticky, in which we find that even high values of $\varepsilon$ can lead to a performance higher than that of SS. For $\varepsilon = 0.5$ and $\varepsilon = 0.75$ we can compare with the results for \EG and find much higher throughput. This shows that the sticky counter has a similar effect than the value of $\varepsilon$, as it decreases the number of STAs reassociating at any given point in time. Thus allowing most STAs to learn properly from the network. We can also observe that for $\varepsilon = 0.1$ and $\varepsilon = 0.25$ we get a higher slope in the first association rounds than with $\varepsilon = 0.05$. In this case then it seems that minimizing the $\varepsilon$ is counterproductive.

Figure \ref{stickycoplot} shows the comparison of \ES with $\varepsilon = 0.1$ and different values of the sticky counter. While the results are fairly similar, increasing the counter to higher values decreases performance, while keeping it to 1 can prevent us from achieving the best performance. Figure \ref{stickycobox} shows the boxplot of the last iteration for each SC value, in which we find that SC$ = $ 2 keeps the 25th percentile the highest, while for values like 6 and 10 the throughput  decreases.

For \EG the optimal value of $\varepsilon$ is $0.05$, and for \ES it is better to keep it at $0.1$. The sticky counter should be 2 for optimal performance. In the case of \ES however we can infer that the value of $\varepsilon$ is less significant, as the stickiness can compensate for a higher $\varepsilon$. It can also be observed that the lower $\varepsilon$ values create a steeper slope of learning (i.e., we learn faster). Also, it bears mentioning that using a low $\varepsilon$ value makes the algorithm atemporal, as the decreasing $\varepsilon$ would require an additional mechanism to detect changes in the network to reset the initial value. 


\section{A reality check for \ES}\label{perfEval2}

Once we have characterized the gains of \ES compared to \EG in the previous section, we focus now on the performance of \ES under more realistic conditions, such as when not all STAs are agent enabled, the STAs arrive progressively, or move around. Also, we discuss the performance of our MAB enabled solution when compared with a traditional load-aware AP selection mechanism.


\subsection{Not all STAs are 'agent' enabled}

\begin{figure}[th!!!!!!]
    \centering
    \includegraphics[width = .46\textwidth]{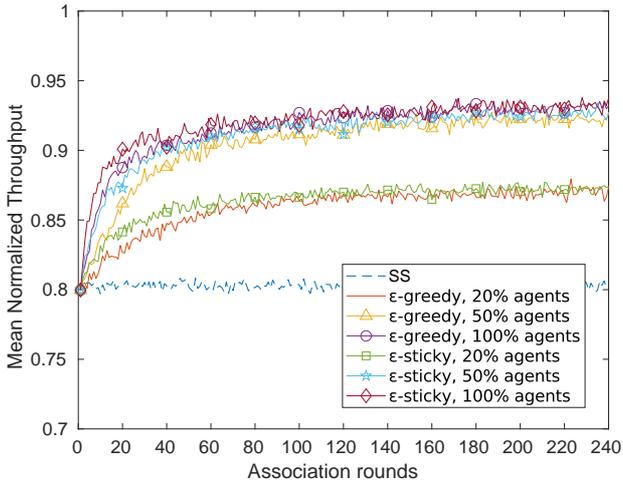}
    \caption{Effect of the fraction of STAs with an active agent on the mean normalized throughput.}
    \label{AgentEvol}
\end{figure}

In this Section we want to test if the $\varepsilon$ algorithms are effective when there are STAs not implementing AP selection agents. Here, we will test networks with different amounts of stations with agents, and find how much the density of stations with agents affects performance. We will use the same scenario as in the previous section, although with the optimal $\varepsilon$ values obtained there for the results.

Figure \ref{AgentEvol} shows the effect of having 20\%, 50\% or all STAs equipped with an agent. We find that even with a low amount of agents we can obtain better performance than in the case where all STAs use the SS method. 

With 20\% of agents, the average throughput increases from 0.8012 with SS to 0.8736 and 0.8741 for \EG and \ES, respectively. When the 50\% of the STAs have an agent, we achieve a throughput of 0.9226 and 0.937. Note that these results are almost identical to those achieved with 100\% of STAs having an agent.

The only difference between implementing agents in 50\% or 100\% of the STAs can be observed in the first 40 association rounds, where their respective learning curves have different slopes. More agents leads to a higher slope and faster learning. We can also observe that with \ES we always have a higher slope than with $\varepsilon$-greedy.

\begin{figure}[ht]
    \centering
    \includegraphics[width = .44\textwidth]{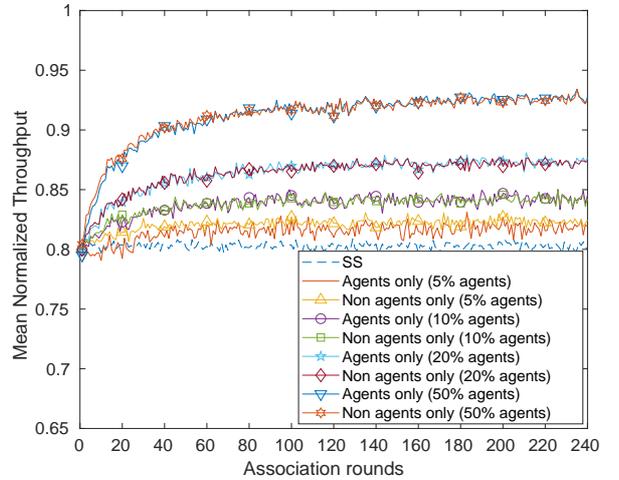}
    \caption{Mean normalized throughput of agent STAs vs. non agent STAs. }
    \label{AgvsNags}
    \end{figure}

From Figure \ref{AgentEvol}, it cannot be deduced if the observed throughput gains are shared between agent and non-agent enabled stations. This comparison can be found in Figure \ref{AgvsNags}. It can be observed that both type of stations benefit from the presence of agent-enabled stations in the same proportion. Indeed, only slight differences in the average throughput between both type of stations are observed for the case where only the 5\% of STAs are equipped with agents, and only during the initial association rounds. Interestingly, the most benefited stations are those that do not equip an agent, as they avoid exploring APs that result in a poorer experience. 

\begin{figure}[ht]
    \centering
    \includegraphics[width=0.44\textwidth]{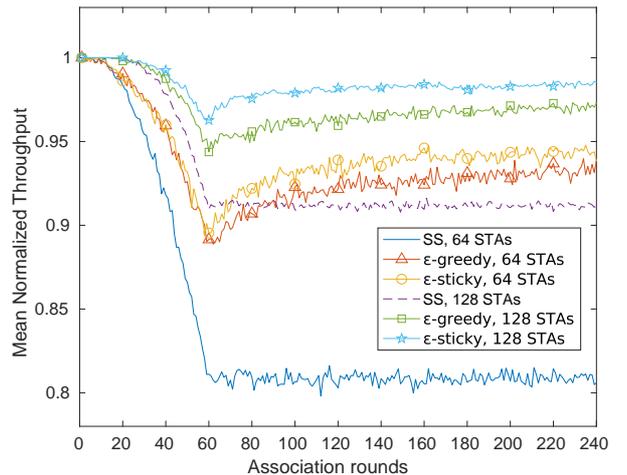}
    \caption{Temporal evolution of the mean normalized throughput when STAs appear progressively.}
    \label{Fstart}
\end{figure}

In summary, we have shown that the network performance improves even if few STAs are equipped with an agent. Moreover, we have also shown that non agent-enabled STAs also benefit from the presence of agent-enabled stations, as the improvement on the use of spectrum resources is shared among all active players.


\subsection{Progressive arrival of STAs}

In this section, we want to consider a non-stationary case with respect to the number of active STAs in the network by allowing STAs to arrive uniformly at random during the first 60 association rounds (3 hours). All STAs will begin their requests in an association round between 1 and 60. We will consider the 64 STAs and the 128 STAs scenarios.

\begin{figure*}[htb]
\centering
   \begin{subfigure}[b]{.44\textwidth}
        \includegraphics[width=\textwidth]{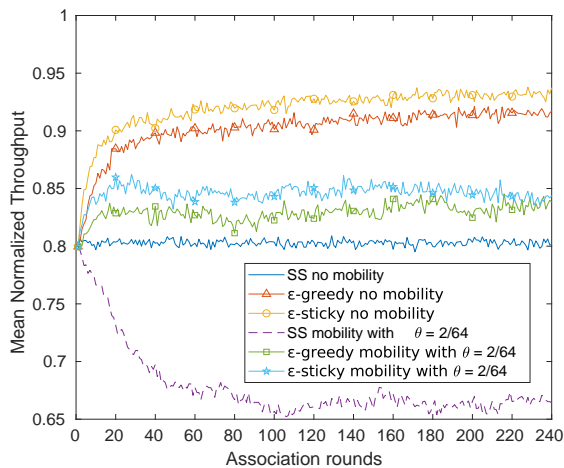}
    \caption{Temporal evolution of the mean normalized throughput when 2 STAs move each round.}
    \label{Mob1}
    \end{subfigure}
   \begin{subfigure}[b]{.455\textwidth}
        \includegraphics[width=\textwidth]{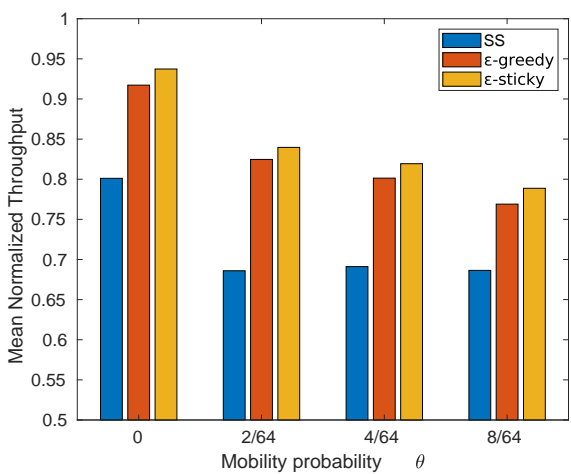}
    \caption{Mean throughput for the last association round as mobility increases.}
    \label{mobcomp}
    \end{subfigure}
     \caption{Impact of STA mobility on throughput achieved.}
     \label{Incs}
\end{figure*}

We show the throughput evolution in Figure \ref{Fstart}, where we find that both algorithms are effectively learning during this initial phase and avoid ever reaching the low values of the SS method. In this initial phase where the STAs appear, for 64 STAs we can observe that the \ES  algorithm performs the same as \EG until all STAs are already in the network, at which point it starts to learn faster than $\varepsilon$-greedy. By the last round, \EG achieves a 15.72\% increase in the throughput over SS, and \ES improves this to 17.24\%. In terms of their reassociation ratios, for each \ES reassociation, \EG performed 4.88 ones.

For 128 STAs however, we find that \ES learns faster and manages to avoid the decrease in throughput observable with $\varepsilon$-greedy. On the last round, \EG and \ES achieve a throughput increase over SS of 6.84\% and 8.20\% respectively, with the reassociation ratio being 2.53 between \EG and $\varepsilon$-sticky.

With the added complexity of the STAs initializing at different intervals, both \EG and \ES not only still outperform the SS method for both the transient phase and the stationary phase, but they also react as the STAs appear, keeping the network on a higher average throughput every step of the way. In both cases, \ES also manages to obtain better performance than \EG with less than half the reassociations.

\subsection{STA mobility}

This section studies the effect of STA mobility on the network throughput and the capacity of our algorithms to cope with it. 

We continue to study the clustered environment of 16 APs and 64 STAs with a random load of mean 4 Mbps per STA, with $\varepsilon = 0.05$ for \EG and $\varepsilon = 0.1$ for $\varepsilon$-sticky. Mobility in our simulations works so as to conserve the clustered environment. The STAs can choose to move to any of the existing clusters in the scenario uniformly at random. Once they have chosen a cluster, the new position inside the cluster is also chosen uniformly at random. At each association round, STAs may change location with a given \textit{mobility probability $\theta$}. 

Note that the standard association mechanism keeps the STA on the same AP for as long as the RSSI is higher than the CCA. Once the signal from the AP is lost, a full scan is performed and the STA selects a new AP with the strongest signal available. We follow the same approach for all STAs, agent and non-agent enabled. Moreover, for \EG and \ES, any STA movement makes the rewards obtained previously useless. Therefore, we considered that agent-enabled STAs reset their reward when they detect a change on the environment (i.e., the APs they observe, and/or the RSSI level from them), meaning that the learning restarts with every new position.

Figure \ref{Mob1} shows the throughput  evolution with and without mobility. For this case we use a mobility probability $\theta = \frac{2}{64} = 0.03125$ that allows, on average, that 2 STAs move at each association round. After moving to another cluster, STAs that still have their previous AP in range will maintain their previous association, and this leads to the decrease of throughput over time observable in the SS case. For the \EG and \ES algorithms this is not such a serious issue, as their exploration allows them to leave their initial AP, and once they have acquired enough rewards to model the new network configuration, they will exploit or stick to whichever AP offers the best performance.

Both \EG and \ES reach higher throughput than the default association, as well as keep it stable. If we compare with the static placement, we can observe that the performance gain of \EG and \ES over SS is higher when there is movement. For the last association round, \EG and \ES show an increase of 26.1\% and 27\% over SS when STAs relocate, while in the static case, we find a 14.49\% and 17\% increase respectively. In the first rounds we can observe that the trend for \EG and \ES is the same in both cases, with \ES learning faster than $\varepsilon$-greedy.

Figure \ref{mobcomp} shows the average throughput achieved in the last association round for different movement probabilities, starting with the static placement and going up to (in average) 8 STAs being re-located per round, a 12.5\% movement probability. Here we can observe that the \EG and \ES performance decreases as more STAs move. This is a result of the algorithms having less time to learn the network state, i.e., the lower the mobility probability, the higher number of association rounds that the MAB-based algorithms have for learning. Still, in all cases we can find that both \EG and \ES outperform SS, with \ES reaching higher throughput than $\varepsilon$-greedy. 

In this section we have shown that the SS mechanism struggles with STA mobility, but both \EG and \ES can cope with it thanks to exploring, as they avoid staying on APs that offer poor performance. We have also found that these MAB algorithms require some environment stability to be able to properly work. Here, the faster learning curve of \ES is beneficial, as the algorithm takes less association rounds to obtain a favourable association for the STA.

\subsection{MABs-based vs a load-aware AP selection mechanism}

We have chosen to use Reinforcement Learning to improve on the AP selection mechanism, but this can also be achieved through other methods, such as using a specific protocol or mechanism. In this section we compare the performance of our MAB-based algorithms to a load-aware AP selection mechanism. Then, the aim of this section is to observe if our MAB-based solutions can offer a better performance, especially by solving situations where the load-aware mechanism fails.

\begin{figure}[htb]
    \centering
    \includegraphics[width=0.5\textwidth]{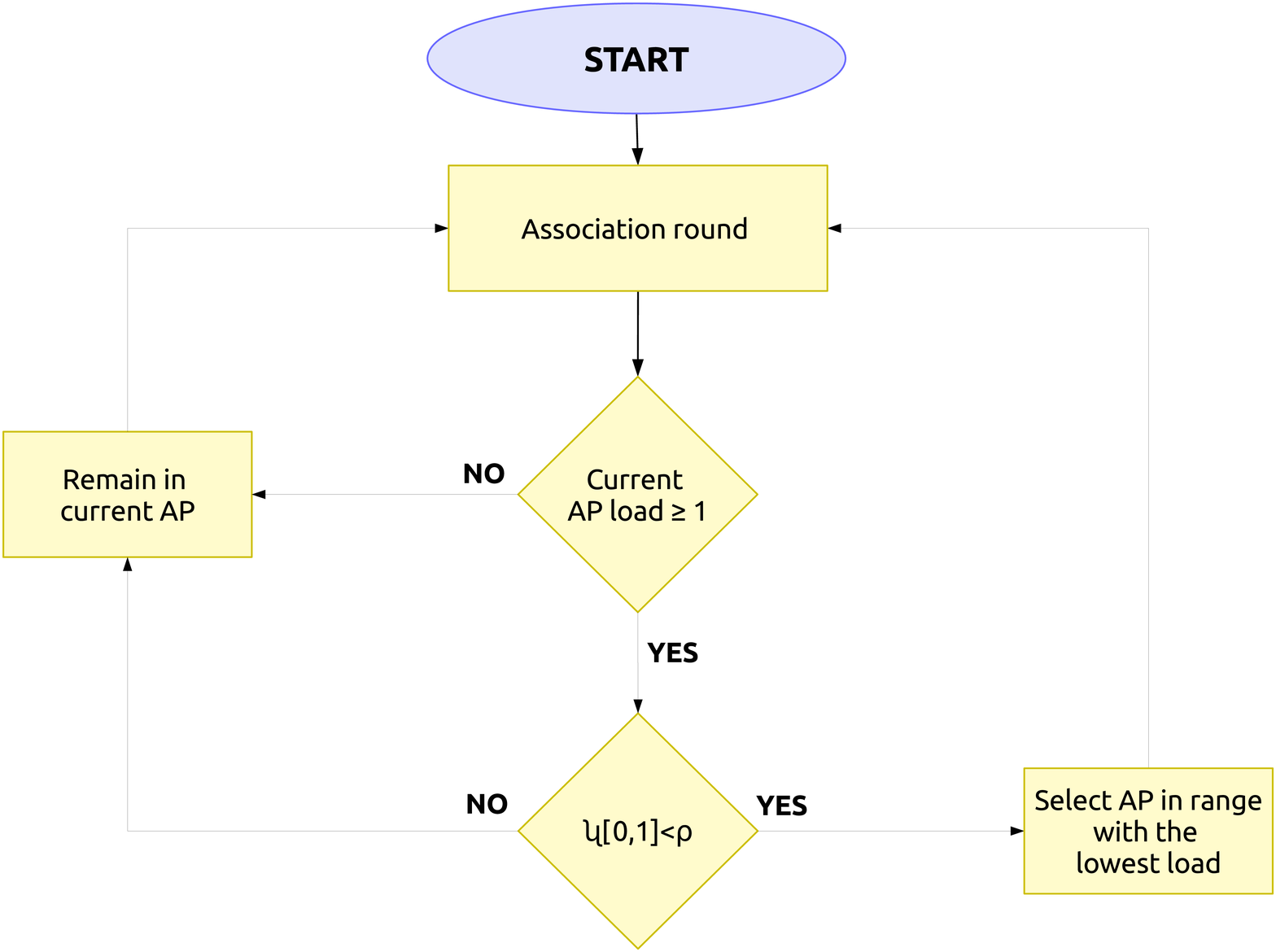}
    \caption{Load-aware decision process.}
    \label{fh}
\end{figure}

The load-aware AP selection mechanism considers APs periodically broadcast their current traffic load (in bps) in their beacons. Then, those STAs that are not satisfied with current association may decide to remain associated to the current AP, or to reassociate to a new AP with a certain \textit{reassociation probability} $\rho$. If they decide to reassociate, they choose the AP with the lowest instantaneous load. The decision process can be found in Figure \ref{fh}.\footnote{In the simulations, at each association round, we randomize the order STAs take decisions to avoid sequential decisions affect the obtained results.}

We compare all three mechanisms (i.e., \EG, \ES and load-aware AP selection mechanism) in the scenario with 16 APs and 64 STAs, and in the following three cases: a) no mobility, and a  constant traffic load of 4 Mbps; b) no mobility, and a variable traffic load of mean 4 Mbps (i.e., the load of each STA may change at each iteration uniformly at random from 1 to 7 Mbps); and c) mobility (12.5 \%), and a variable traffic load of mean 4 Mbps.

\begin{figure*}[ht]
\centering
   \begin{subfigure}[b]{.329\textwidth}
        \includegraphics[width=\textwidth]{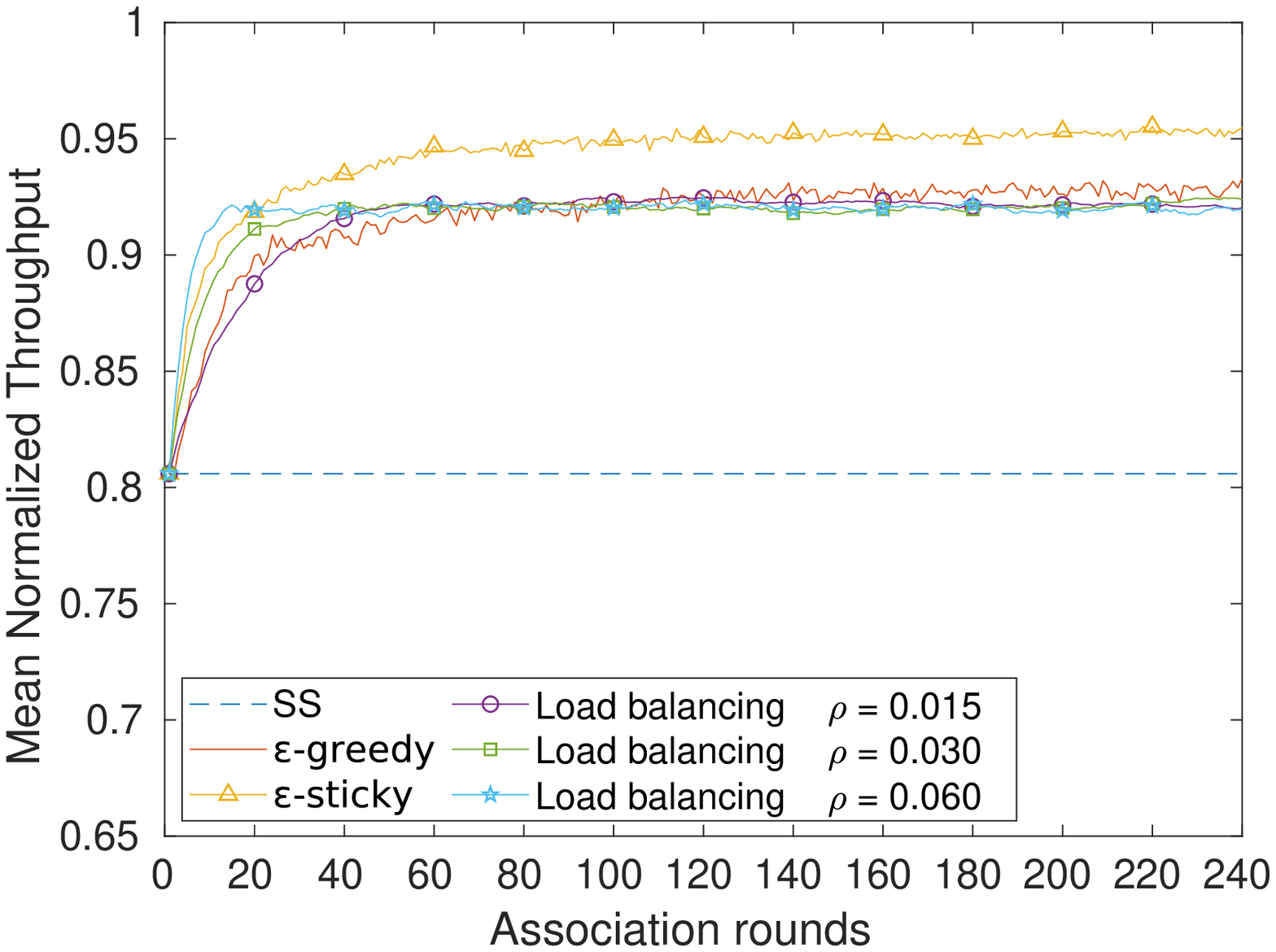}
    \caption{Normalized throughput for static load of 4 Mbps. }
    \label{HeurStat}
    \end{subfigure}
     \begin{subfigure}[b]{.329\textwidth}
        \includegraphics[width=\textwidth]{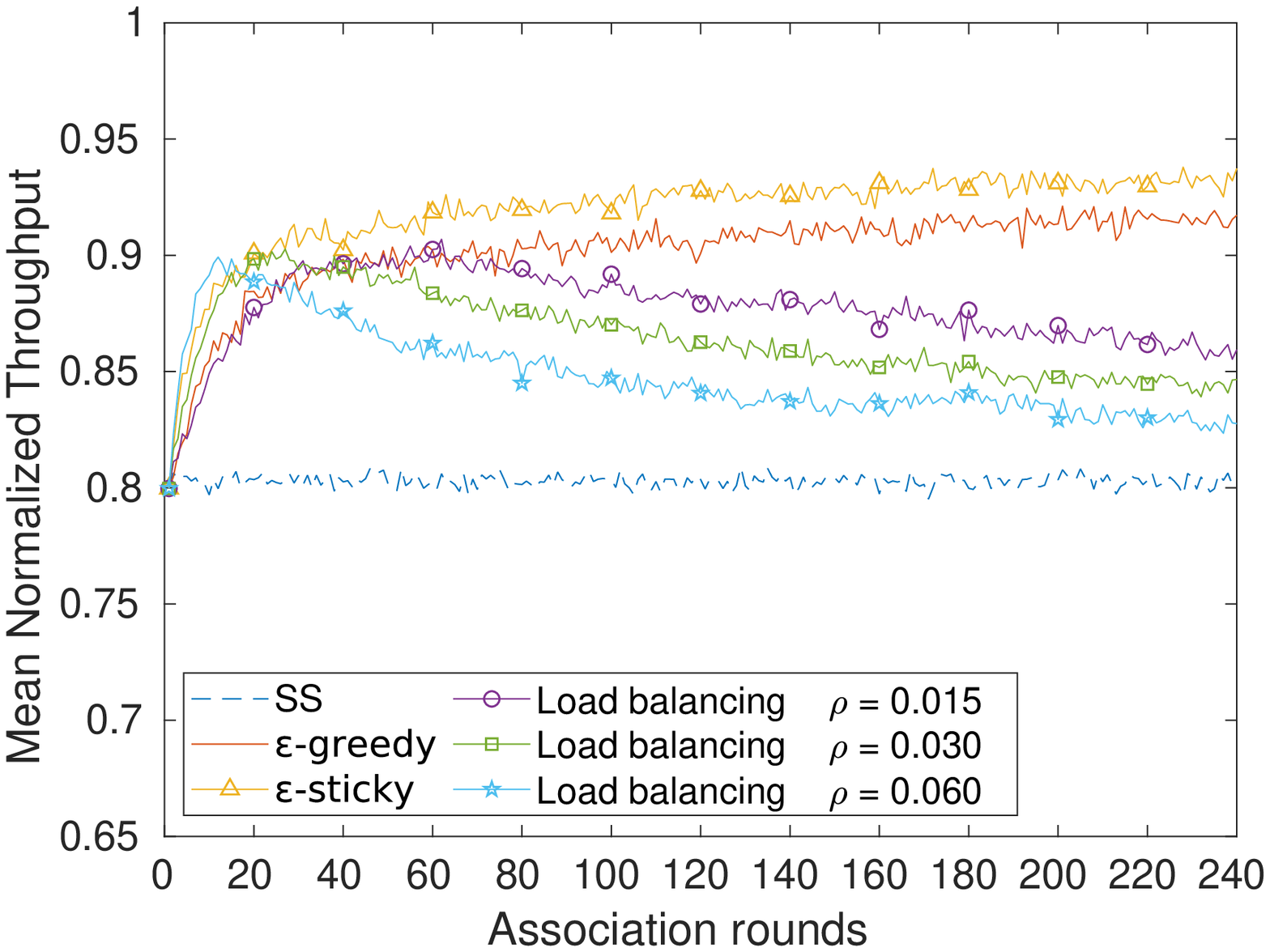}
    \caption{Normalized throughput for random load with mean 4 Mbps.}
    \label{HeurRand}
    \end{subfigure}
    \begin{subfigure}[b]{.329\textwidth}
        \includegraphics[width=\textwidth]{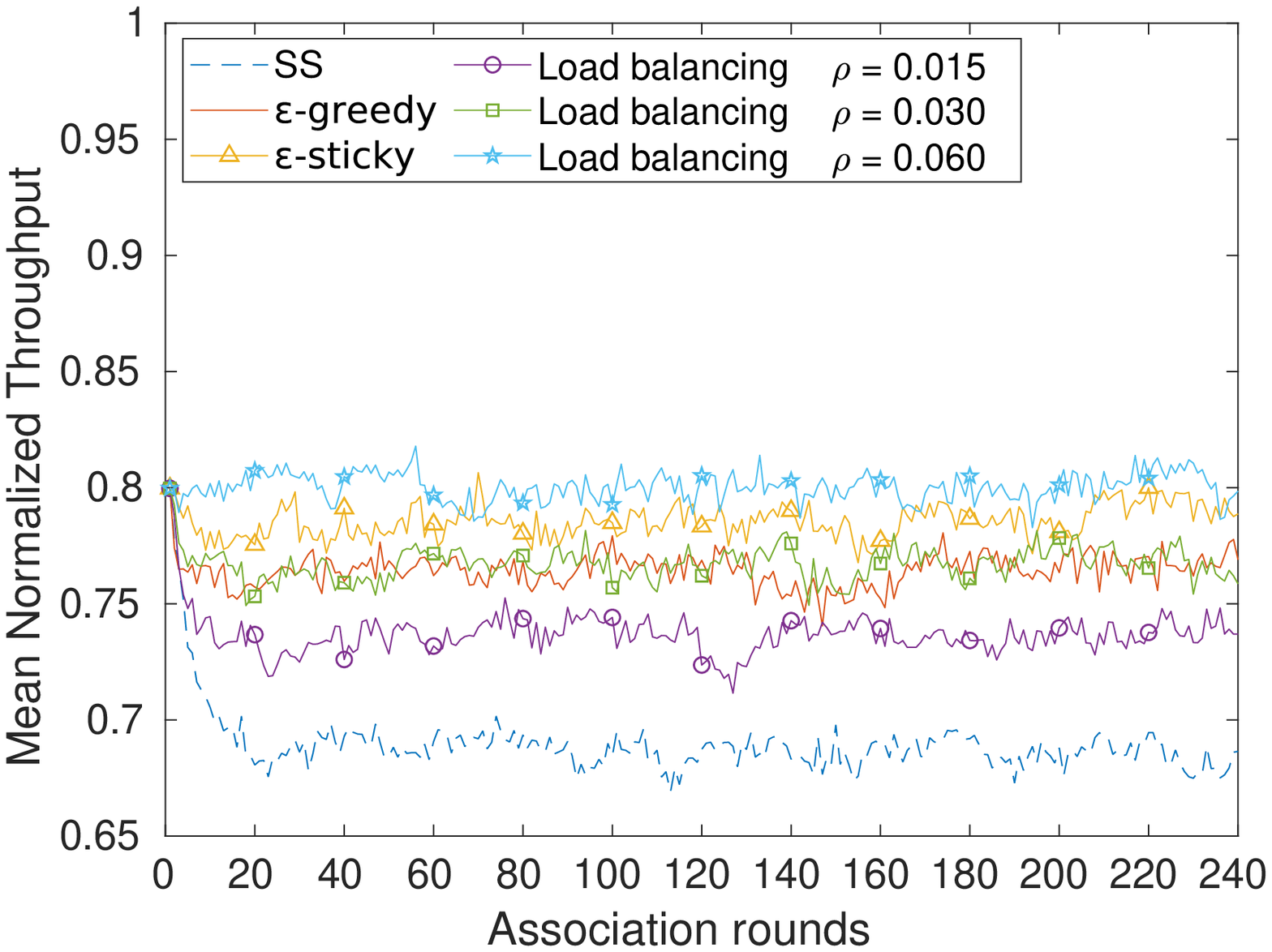}
    \caption{Normalized throughput for random load with mean 4 Mbps and mobility.}
    \label{HeurRandMob}
    \end{subfigure}
     \caption{Throughput over time. }
     \label{Heuristics}
\end{figure*}

Figure \ref{HeurStat} shows the throughput over time for the case with no mobility and a constant load. We can observe that the load-aware mechanism obtains a similar performance than the \EG algorithm for all reassociation probabilities. The only difference can be found in the convergence rate for the first association rounds, which is faster for high reassociation probabilities. We can also observe that there is less fluctuation in the throughput for the load-aware as it avoids exploration. Lastly, we can also observe how \ES beats all other methods when constant traffic loads are considered. The reason for this is that \ES relies on the past rewards  to rank the rest of the APs, which allows it to always select the best possible AP even if it does not fully satisfy the STA.

Figure \ref{HeurRand} shows the case with no mobility and variable traffic loads. We can observe that the load-aware starts achieving a similar throughput than the \EG algorithm, but the throughput starts to decrease over time. This is likely due to reassociation decisions are made with data from the last round only, as opposed to the \EG and \ES algorithms, which use previous knowledge. In this particular case, the load-aware mechanism performs worse for higher reassociation probabilities, as more STAs change to APs that do not offer the expected good performance. 

Finally, in Figure \ref{HeurRandMob}, we show the performance with variable loads and mobility. Here, we observe that the load-aware mechanism performs worse than \EG and \ES with a reassociation probability of 1.5\%, similar to \EG when the probability is of 3\% and better than \ES when it is of 6\%. A high mobility rate reduces the chances for the \EG and \ES algorithms to find a good association for the STAs. It is because they reset the accumulated rewards almost every time they move, and so they do not get a chance to gather enough knowledge from the surrounding APs. On the contrary, the load-aware mechanism reacts faster to the situation found at each round, leading to a better performance in these highly chaotic situations.  Both \EG and \ES could be further enhanced to retain information from past configurations, including their location, so that the gathered knowledge would not be lost and can be used in the future.

 In summary, the use of MABs for AP selection, alone, or in combination with other protocols, is promising for three reasons: i) the use of rewards that consider previous experience to characterize the network response,  ii) the rewards are updated constantly, capturing and adapting to changes in the environment, and iii) The random exploration of these algorithms, thus, avoiding deterministic actions, such as when all STAs in a similar situation make the same choice at the same time, which results in a low performance for all of them. In the load-aware mechanism, two STAs in similar positions and conditions will always make the same choice. The default association mechanism does this as well by always relying on the RSSI. By adding randomness into the mechanism, MABs avoid these cases, thus the two STAs would select different APs, avoiding overcrowding an AP.


\section{Related Work}\label{RelWork}

AP selection and load balancing have been extensively studied as a way to improve network throughput. A scheme is proposed in \cite{1313270} where neighboring APs compare their traffic loads to decide if they should force the disassociation of a user so that it reassociates to an underloaded AP. The authors in \cite{4151354} use the delay between a \textit{probe request} being sent and a \textit{probe response} being received as a measure of the load of the AP, and base their association scheme on picking the AP with the lowest delay instead of the lowest RSSI. In \cite{4796200} the authors use cell breathing techniques to balance the load among APs by modifying the transmission power of the \textit{beacons} sent by the AP, virtually reducing their coverage area so that STAs reassociate to other uncongested APs. A solution based on inter-AP interference is proposed in \cite{7876094}, where the STAs estimate the Signal to Interference plus Noise Ratio (SINR) from interfering APs by sending \textit{probe requests} to all APs. Then, from the \textit{probe responses} received, they can estimate the SINR, and choose the best one to find the optimal association for each STA.

In \cite{6181180}, the authors propose the use of a decentralized neural network with a single hidden layer that uses the SNR, number of STAs detected, probability of retransmissions and channel occupancy as inputs to predict the throughput achievable for each AP in the network, as well as the  optimal association to the one that maximizes it. To the best of our knowledge there are no other papers in the area of RL applied to user association. The use of MABs however is starting to be familiar to solve optimization problems in decentralized and complex scenarios. For example, the authors in \cite{7792374} give an overview of the multi armed bandits problem, as well as its applications in wireless networks as a way to solve resource allocation issues. The work in \cite{wilhelmi2019collaborative,wilhelmi2019potential} uses several Reinforcement Learning algorithms to find the optimal selection of channel and transmission power for each AP in a network. In \cite{6939716} the authors use MABs in device to device communication systems to help users choose the optimal channel and improve their performance. 


\section{Conclusions}\label{concl}

In this paper we have tackled the problem of finding a feasible association in Enterprise WLANs in a decentralized way. We have used MAB algorithms to give the STAs the ability to explore their association options and find a suitable AP-STA pairing. We have extended the \EG algorithm by adding stickiness to it, which can greatly reduce the amount of reassociations needed to find a solution that satisfies most of the active STAs.

We have tested the suitability of these algorithms for different APs and STAs distributions, showing that they can always find a better AP-STA configuration than the Strongest Signal method. We have also investigated how \EG and \ES algorithms can cope with variable STA loads, number of channels and channel bandwidths. When comparing both algorithms, we show that while \EG can perform well in terms of throughput, it is inefficient in terms of reassociations. With \ES we not only achieve higher throughput values, but we also reduce the amount of reassociations required to find a suitable AP-STA configuration.

We have also studied non-stationary cases: in the first one where the STAs appear progressively, both \EG and \ES can mitigate the effect of new STAs, and keep the throughput high. In the second case we limit the amount of STAs that have an agent implementing the proposed algorithms. We find that even a low number of agents can increase the performance of the whole network. Indeed, we observe that when half of the STAs implement an agent, the system performance is close to that of when all STAs are equipped with agents. Finally, we have compared our MAB based approach to a load-aware AP selection mechanism, and highlighted the advantages and disadvantages of both methods.

Future challenges include to fully assess the impact of different degrees of mobility in the performance of the proposed schemes, and how it can be handled from the agent perspective when the STA moves between different locations. We have done a limited study in which we reset the rewards after a movement is detected, but this can be clearly improved further by keeping location history and not discarding past rewards. It would also be interesting to design hybrid schemes by extending load-aware like methods with exploration-exploitation capabilities from MABs to benefit from both perspectives, and so further strengthen their advantages. We, indeed, think this is the path to follow to successfully introduce RL solutions in  networking.

\section*{Acknowledgements}\label{ACKs}

This work has been partially supported by a Gift from CISCO University Research Program (CG\#890107) \& Silicon Valley Community Foundation, by the Spanish Ministry of Economy and Competitiveness under the Maria de Maeztu Units of Excellence Programme (MDM-2015-0502), by WINDMAL PGC2018-099959-B-I00 (MCIU/AEI/FEDER,UE), and by the Catalan Government under grant SGR-2017-1188.
\

\section*{Copyright}
\textcopyright 2020. This manuscript version is made available under the CC-BY-NC-ND 4.0 license.

\appendix
\section{Airtime calculation}\label{appA}

Here we give a more detailed look at the airtime calculation for the STAs. The transmission time for a frame is calculated as:

\begin{align}
     T(r_i,r_{L,i}) = T_{\text{data}}(L,r_i) + \text{SIFS} + T_{\text{ack}} +  \text{DIFS} + T_e
\end{align}
where
\begin{align}
    T_{\text{data}}(r_i) = T_{\text{PHY-HE-SU}} +  \bigg\lceil\frac{L_{\text{SF}} + L_{\text{MH}} + L_i + L_{\text{TB}}}{r_i} \bigg\rceil \sigma
\end{align}
and
\begin{align}
     T_{\text{ack}}(r_{L,i}) =  T_{\text{PHY-legacy}}  + \bigg \lceil \frac{L_{\text{SF}} + L_{\text{ACK}}+ L_{\text{TB}}}{r_{L,i}} \bigg \rceil    \sigma_{\text{legacy}}
\end{align} 

The final airtime required is calculated as stated in equation \ref{eq:flow_util} in Section \ref{appref}.

\bibliographystyle{unsrt}
\bibliography{Bib.bib}

\end{document}